\documentclass[epjST,final]{svjour}

\usepackage{graphicx}
\usepackage{amsmath,amssymb,amsfonts}
\usepackage{bbm}
\usepackage{epstopdf}
\usepackage{braket}
\usepackage{color}
\newcommand{\eV}{\ \mathrm{eV}}

\newcommand{\msu}{\uparrow}			
\newcommand{\msd}{\downarrow}		
\newcommand{\real}{\operatorname{Re}}
\newcommand{\imag}{\operatorname{Im}}
\newcommand{\bighat}[1]{\stackrel{\vartriangle}{#1}}
\DeclareMathOperator{\Tr}{Tr}

\definecolor{rosso}{rgb}{0.8,0.0,0.0}

\begin{document}
\title{Realistic theory of electronic correlations in nanoscopic systems}

\author{Malte Sch\"uler\inst{1} \and Stefan Barthel\inst{1} \and Tim Wehling\inst{1} \and Michael Karolak\inst{2} \and Angelo Valli\inst{3} \and Giorgio Sangiovanni\inst{2}}
%
\institute{Institut f\"ur Theoretische Physik, Universit\"at Bremen, Germany \and Institut f\"ur Theoretische Physik und Astrophysik, Universit\"at W\"urzburg, Germany \and Istituto Officina dei Materiali (CNR-IOM) and Scuola Internazionale Superiore di Studi Avanzati (SISSA), Via Bonomea 265, 34136 Trieste, Italy}
\abstract{
Nanostructures with open shell transition metal or molecular constituents host often strong electronic correlations and are highly sensitive to atomistic material details. This tutorial review discusses method developments and applications of theoretical approaches for the realistic description of the electronic and magnetic properties of nanostructures with correlated electrons. First, the implementation of a flexible interface between density functional theory and a variant of dynamical mean field theory (DMFT) highly suitable for the simulation of complex correlated structures is explained and illustrated. On the DMFT side, this interface is largely based on recent developments of quantum Monte Carlo and exact diagonalization techniques allowing for efficient descriptions of general four fermion Coulomb interactions, reduced symmetries and spin-orbit coupling, which are explained here.
With the examples of the Cr (001) surfaces, magnetic adatoms, and molecular systems it is shown how the interplay of Hubbard $U$ and Hund's $J$ determines charge and spin fluctuations and how these interactions drive different sorts of correlation effects in nanosystems.
Non-local interactions and correlations present a particular challenge for the theory of low dimensional systems. We present our method developments addressing these two challenges, i.e. advancements of the dynamical vertex approximation and a combination of the constrained random phase approximation with continuum medium theories. We demonstrate how non-local interaction and correlation phenomena are controlled not only by dimensionality but also by coupling to the environment which is typically important for determining the physics of nanosystems.
} 
\maketitle
\section{Introduction}
\label{intro}

The electronic structure problem in nanostructures involves two sources of complexity: 
First, on a single-particle level, a large Hilbert space can be required to describe 
how electronic orbitals adjust to a complex arrangement of many inequivalent atoms, 
which by definition of a nanostructure lacks crystal translational symmetry in one or more spatial directions. 
Second, electron-electron interactions are usually pronounced in systems with reduced dimensionality, 
triggering strong electronic correlation effects.

In this review, we report on recent advances on realistic material simulations of nanostructures 
hosting electronic correlations which were made within the DFG Research Unit FOR 1346. 
Our strategy, here, is to combine first-principles methods 
--- most prominently density functional theory (DFT) and the GW approximation --- with approaches for the description 
of strong electronic correlations \cite{lda++,anisimov_first-principles_1997,amadon_plane-wave_2008},  
in particular dynamical mean-field theory (DMFT) \cite{georgesRMP68} and related approaches/extensions thereof. 
The general idea is to account for atomistic details of the nanosystems based on the first-principles methods, 
which are then used to derive realistic model Hamiltonians describing the electron correlation 
in effective low energy Fock spaces constructed out of a reduced set of single-particle orbitals. 
In this way, realistic and complex nanostructures are in principle within reach for approaches like DMFT, 
which critically rely on Hilbert space dimensions not being too large.

The review is structured as follows: In section \ref{sec:interface} we explain how 
first-principles and correlated electron approaches can be interfaced in such a way that 
treatments of complex (nano)structures becomes possible 
and how the resulting many-body problems can be solved efficiently. 
Section \ref{sec:impsolv} is dedicated to a brief description of the impurity solvers used by us in this context. 
In Section \ref{sec:applications} we discuss applications of this approach to so-called ``Hund's impurity systems'', 
molecular systems as well as the Cr (001) surface. 
Section \ref{sec:non_loc} explains a novel approach for the derivation of realistically screened interaction terms 
in complex nanosystems and discusses progress on how to treat non-local interactions and non-local correlations in nanosystems.

\section{Interface of density functional and dynamical mean field theory for the simulation of complex structures}
\label{sec:interface}

Considerable efforts have been focussed on the modelling of correlated nanoscale systems using DMFT techniques, 
e.g. \cite{Potthoff_Nolting_PRB99,tanskovicPRL91,Florens_PRL07,jacobPRL103,Jacob_Kotliar_PRB2010,valliPRB86,turkowskiJCP136,kabirJPCM27,leePRB91,davids_review,valliPRB92,valliPRB94}. 
Here, we describe the approach to a material realistic DMFT scheme for nanosystems 
which has been implemented by the authors during the course of FOR 1346, and which is particularly flexible.

The DFT simulation of a material or nanostructure yields an auxiliary model 
of non-interacting electrons in the material 
via the Kohn-Sham wave functions $\{\ket{\mathbf{k},N}\}$ and energies $\varepsilon^N_\mathbf{k}$. 
The first step of any approach combining ab-initio and many-body methods (``DFT++'') 
is to consider this Kohn-Sham system as an effective single-particle starting point for the electronic problem. 
Then, one can identify a correlated subspace $\{\ket{m}\}$, where the Kohn-Sham system 
can be augmented by interactions beyond those included within DFT. 
Here, $m$ is a combined site- and (spin-)orbital-index. 
Let us also stress that, at this stage, the distinction between the correlated orbitals 
and ``the rest'' of the system is just formal: 
beyond the actually correlated orbitals (e.g. 3$d$), what we call correlated subspace 
can indeed contain also orbitals (such as $p$ or $s$) 
on which we are \emph{not} always going to explicitly add a Coulomb interaction, 
but that we nevertheless want to keep in 
our ``low-energy'' basis set and treat as ``active'' degrees of freedom. 
Independently on these formal aspects, the procedure always requires the addition of a double-counting correction 
in order to account for the interactions already present within DFT.
 
\begin{figure}\sidecaption
\includegraphics[width=0.6\columnwidth]{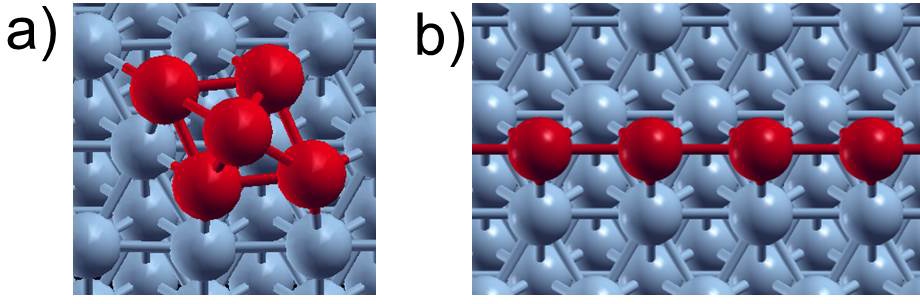}
\caption{Sketch of a non-periodic (panel $a$) and of a periodic (panel $b$) system of correlated (red) atoms hybridized to a substrate (blue atoms). Adapted from Ref.~\cite{khajetoorians_science_2013}. Reprinted with permission from AAAS.} 
\label{fig:nanoPic}
\end{figure}

To make calculations feasible local interactions are often assumed\footnote{A discussion of how appropriate realistic local effective interaction terms can be obtained is given in Section \ref{sec:non_loc}.}. 
The low-energy effective Hamiltonian describes a generalized multi-impurity Anderson Model (AIM), 
which also includes the coupling of the correlated subspace to leads or substrate atoms.
The latter is described by a frequency- and momentum-dependent potential denoted, in Eq.\ref{Green}, 
by the Greek letter $\Gamma$. It represents a \emph{fixed} physical bath, 
i.e. a particle reservoir determined by the geometry of the problem and by the kind of coupling 
between the correlated atoms and the rest of the system. 
This bath is not to be confused with the self-consistent DMFT bath of the auxiliary single impurity problems. 
An example of this is sketched in Fig.~\ref{fig:nanoPic} where the red atoms host the correlated orbitals 
and the blue ones represent the physical bath.
The orbitals $\ket{m}$ spanning the correlated subspace should be chosen as localized as possible 
or atomic-like in order to have a consistent description in terms of AIM or Hubbard-like models with local interactions.

It is useful to introduce a matrix notation in the basis of the (spin-)orbitals of all atoms in the low-energy subspace. 
The ``Pinocchio-hat'' $\vartriangle$ indicates full matrices in this space. 
Since for the DMFT algorithm we sometimes work with sub-blocks of such matrices corresponding to a given atom, 
we indicate these $N_\text{orb} \times N_\text{orb}$ smaller matrices with the standard hat-symbol.
We consider both periodic crystals and inhomogeneous nano-/hetero-structures with reduced, 
or even absent translational symmetry.
For the cases in which the spatial periodicity is (even only partially) preserved, 
the matrix structure reflects the fact that the unit cell can contain more than one atom and each atom contributes, 
in general, more than one (spin-)orbital.
Similarly, in all other non-periodic cases, it describes the real-space structure of all the (multi-orbital) atoms of the nanostructure.
Hence, the $\mathbf{k}$-integrated Green's function can be written as 
\begin{align} \label{Green}
\bighat{G}(i\omega_n) =  
\frac{1}{N_\mathbf{k}} {\sum}_{\mathbf{k}}
\left[ (i\omega_n + \mu)\bighat{1\!\!1} - \bighat{H}(\mathbf{k}) -\bighat{\Gamma}(\mathbf{k},i\omega_n) 
- \bighat{\epsilon}_{\rm DC} - \bighat{\Sigma}({\bf k},i\omega_n)  \right]^{-1},
\end{align}
where $\omega_n = \pi T (2n+1)$ are the Matsubara frequencies at temperature $T$ and $\mu$ is the chemical potential.

The matrix $\bighat{H}({\bf k})$ encodes both inter- and intra-atomic single-particle contributions, 
while $\bighat\Gamma(\mathbf{k},i\omega_n)$ describes the hybridization 
of the orbitals of the correlated subspace to the physical leads or to the substrate. 
In general, $\bighat{\Gamma}(\mathbf{k},i\omega_n)$ and $\bighat{H}(\mathbf{k})$ are obtained via projection of the Kohn-Sham non-interacting Green's function onto the localized states, 
as described in Section \ref{sec:projector}. 
Since these two quantities represent physical properties of the system they are fixed for the rest of the calculation, or recomputed at each iteration of a charge self-consistent DFT+DMFT (see Sec.~\ref{sec:chargeSC}).
If the system is a non-periodic nanostructure 
-- as the one shown in Fig.~\ref{fig:nanoPic}a -- and there is no translation symmetry 
in any spatial direction, $N_\mathbf{k}$ is formally equal to 1, and the sum drops out. 
In that case $\bighat{H}$ and $\bighat{\Gamma}(i\omega_n)$ are $\mathbf{k}$-independent matrices, the latter being the ``standard'' frequency-dependent hybridization function, often also called $\bighat{\Delta}(i\omega_n)$ in the context of the Anderson impurity model.
If instead the correlated atoms form a periodic structure 
-- as e.g. a thin film or a nanowire of a correlated material on a substrate as sketched in Fig. \ref{fig:nanoPic}b -- 
the $\mathbf{k}$-summation extends over the corresponding one- or two-dimensional Brillouin zone. 
In all cases, the double-counting correction $\bighat{\epsilon}_{DC}$ 
retains a matrix structure, in order to account for the presence of more than one orbital/atom. 
The self-energy $\bighat{\Sigma}({\bf k},i\omega_n)$ describes the dynamical effects 
of electron-electron interaction. 
Within DMFT, the self-energy is purely local, i.e., it carries no ${\bf k}$-dependence 
and it is a block-diagonal matrix in the atoms. 
A structure in ${\bf k}$-space for the self-energy, as well as off-diagonal elements 
between different atoms can be obtained by taking into account 
non-local electronic correlations beyond DMFT, as discussed in Section \ref{sec:non_loc}.

The simplest case of a simulation combining ab-initio and many-body approaches is represented by a single correlated adatom or an atom coupled to some leads: The solution of the low-energy model is obtained via a multi-orbital quantum impurity solver which is able to yield the self-energy and hence the two-point Green's function, given the bath described by $\bighat{\Gamma}(i\omega_n)$ and the local-level structure of $\bighat{H}$.
For the cases of a non-periodic nanostructure with more than one atom as well as for those of a periodic arrangement of correlated sites, the calculation requires instead a self-consistent adjustment of the Weiss field, i.e. the ``DMFT'' bath, that describes the effect on the $i$-th atom of the other correlated sites of the system.
This is defined as
\begin{align} \label{Weiss}
\hat{\mathcal{G}}^{-1}_{i,N_\text{orb} \times N_\text{orb}}(i\omega_n) = \left(\left. \bighat{G}(i\omega_n) \right|^{\text{atom}\, i}_{N_\text{orb} \times N_\text{orb}} \right)^{-1} + \hat{\Sigma}^{\text{atom}\, i}_{N_\text{orb} \times N_\text{orb}},
\end{align}
where the matrix to be inverted is the $N_\text{orb} \times N_\text{orb}$ block corresponding to the $i$-th atom of the $\mathbf{k}$-summed Green's function of Eq. \ref{Green}.
The Weiss field defines an auxiliary \emph{single-site} quantum many-body impurity model, which must be solved to get the two-point Green's function of the $i$-th site $\hat{G}_\text{imp $i$}(i\omega_n)$. The corresponding self-energy is obtained via the Dyson equation
\begin{align}
\label{sigma}
\hat{\Sigma}^{\text{atom}\, i}_{N_\text{orb} \times N_\text{orb}} = \hat{\mathcal{G}}^{-1}_{i,N_\text{orb} \times N_\text{orb}}(i\omega_n) - \left( \hat{G}_\text{imp $i$}(i\omega_n) \right)^{-1}
\end{align}
Once the self-energy matrices for all sites in the cluster or in the unit cell are calculated via Eq.~\ref{sigma}, the full matrix $\bighat{\Sigma}(i\omega_n)$ is constructed block-wise and inserted back into Eq.~\ref{Green} to get a new local Green's function for the whole system. The self-consistency loop therefore goes on by calculating Eq.~\ref{Weiss} for each site again, until convergence is reached.

\subsection{General projector and local Hamiltonian formalism}
\label{sec:projector}

The DFT++ scheme explained in the previous section relies on extracting $\bighat H(\mathbf{k})$ and $\bighat \Gamma(\mathbf{k},i\omega_n)$ as entering Eq.~\ref{Green} from ab initio calculations. We achieve this using projection operators $P = \sum_m \ket{m}\bra{m}$ on the correlated subspace, as we explain in the following. To define $P$, in practice, the overlaps $\langle m | \mathbf{k},N\rangle$ of correlated local orbitals and Kohn-Sham wave functions in the DFT code are sufficient and taken e.g. from the PAW projectors implemented in the VASP code or more generally using any kind of Wannier function describing the correlated orbitals.

The DFT simulation yields the Kohn-Sham Hamiltonian $H_\text{{KS}}(\mathbf{k})$ together with the corresponding eigenstates $\{\ket{\mathbf{k},N}\}$ and energies $\varepsilon^N_\mathbf{k}$ and thus also the Kohn-Sham Green function:
\begin{equation}
G_\text{{KS}}(\mathbf{k},i\omega_n) = \left[i\omega_n - H_\text{{KS}}(\mathbf{k})\right]^{-1}= \sum_N | \mathbf{k},N\rangle \left[i\omega_n - \varepsilon^N_\mathbf{k} \right]^{-1}\langle \mathbf{k},N | \label{eq:DefKSGreenfunction}
\end{equation}
$G_\text{{KS}}$ is a matrix in the space of all Kohn Sham states. It is particularly for nano\-structures a much higher dimensional object than the matrices occurring in Eq.~\ref{Green}, which act on the correlated subspace only. The projections of $H_\text{{KS}}$ and $G_\text{{KS}}(\mathbf{k},i\omega_n)$ on the correlated subspace yields all quantities entering Eq.~\ref{Green}:
\begin{equation}
\bighat{H}(\mathbf{k})=P H_\text{{KS}}(\mathbf{k})P
\end{equation}
or element wise
$\bigl( \bighat{H} \bigr)_{mm'}(\mathbf{k})= \sum_N \langle m | \mathbf{k},N\rangle \langle \mathbf{k},N |m' \rangle  \varepsilon^N_\mathbf{k}$.
The projection of the Kohn-Sham Green function on the correlated subspace
$$ P G_\text{{KS}}(\mathbf{k},i\omega_n)P $$
then yields the frequency- and momentum-dependent potential entering Eq.~\ref{Green} via
\begin{equation}
\bighat\Gamma(\mathbf{k},i \omega_n)=(i\omega_n + \mu)\bighat{\mathbbm{1}} - \bighat{H}(\mathbf{k})-\left(P G_\text{{KS}}(\mathbf{k},i\omega_n)P\right)^{-1}.
\label{eq:DefDelta}
\end{equation}
In the case of an adatom or a molecule absorbed on a surface (c.f. section \ref{subsec:Hund}-\ref{subsec:molecules}) there is no $\mathbf{k}$-dependence in Eq. (\ref{eq:DefDelta}). Then, $\left(P G_\text{{KS}}(\mathbf{k},i\omega_n)P\right)$ plays the role of the Kohn-Sham impurity Green function, $\bighat{H}$ corresponds to the crystal field and $\bighat\Gamma$ is often also labelled hybridization function $\Delta$. See e.g. Eq. (12) in Ref. \cite{karolak_general_2011}. 
Eq. (\ref{eq:DefDelta}) is often expressed explicitly in terms of hopping matrix elements between impurity orbitals and bath states. If the $\mathbf{k}$-dependence is absent, this takes the following form: $\left(\bighat\Gamma(i\omega_n)\right)_{ii'}=\sum_\nu\frac{V_{i\nu}V_{\nu i'}}{i\omega_n-\epsilon_\nu}$, where $V_{i,\nu}$ connects the impurity orbital $i$ with the bath state $\nu$ at energy $\epsilon_\nu$. The imaginary part of the hybridization function $\bighat\Gamma$ is then simply related to the bath density of states $\rho(\omega)$ if we assume that the hopping $V$ between all impurity orbitals and the bath is the same: $\Im\left(\bighat\Gamma(\omega+i0^+)\right)_{ii}=-\pi|V|^2\rho(\omega)$.

\subsection{Charge self-consistency}
\label{sec:chargeSC}

The DFT++ approach implicitly includes interactions between electrons in the correlated subspace and the rest of the system through the Hartree as well as the exchange correlation potential from DFT. As soon as the many-body part of DFT++ redistributes electrons between correlated and uncorrelated orbitals or also between different sites there will be associated Hartree (as well as possible exchange or correlation) energies and the DFT++ Hamiltonian should be correspondingly updated. In general, it is obviously problematic to obtain the update of the DFT++ Hamiltonian simply from a double-counting correction applied to the correlated subspace only. This can be better achieved by including self-consistency over the charge-density in the DFT++ approach which obviously requires us to work in the full Kohn-Sham Hilbert space of the DFT part of our approach. We calculate the updated electron density of the DFT++ system,
\begin{align}
n(r) = \frac{1}{\beta} \sum_{\mathbf{k},N,N',n} \langle r| \mathbf{k},N\rangle G_{N N'}(\mathbf{k},i\omega_n) \langle \mathbf{k},N' | r \rangle
\end{align}
where $G_{N N'}(\mathbf{k},i\omega_n)=\langle\mathbf{k},N|G(\mathbf{k},i\omega_n)|\mathbf{k},N'\rangle$ is the interacting Green's function of the full system and acts on the same Hilbert space as Eq. (\ref{eq:DefKSGreenfunction}). It includes corrections due to dynamic self-energy effects within the correlated subspace according to
\begin{equation}
G(\mathbf{k},i\omega_n) = \left[i\omega_n - H_\text{{KS}}(\mathbf{k})- \epsilon_{\rm DC} - \Sigma({\bf k},i\omega_n)\right]^{-1}.
\end{equation}
$\epsilon_{\rm DC}$ and $\Sigma$ are quantities defined in the full Hilbert space and are obtained from the corresponding ones ($\bighat{\epsilon}_{\rm DC}$ and $\bighat{\Sigma}$) using the overlaps $\langle m | \mathbf{k},N\rangle$, as explained in Ref.~\cite{amadon_plane-wave_2008}.

\begin{figure}
\centering
\includegraphics[width=0.85\columnwidth]{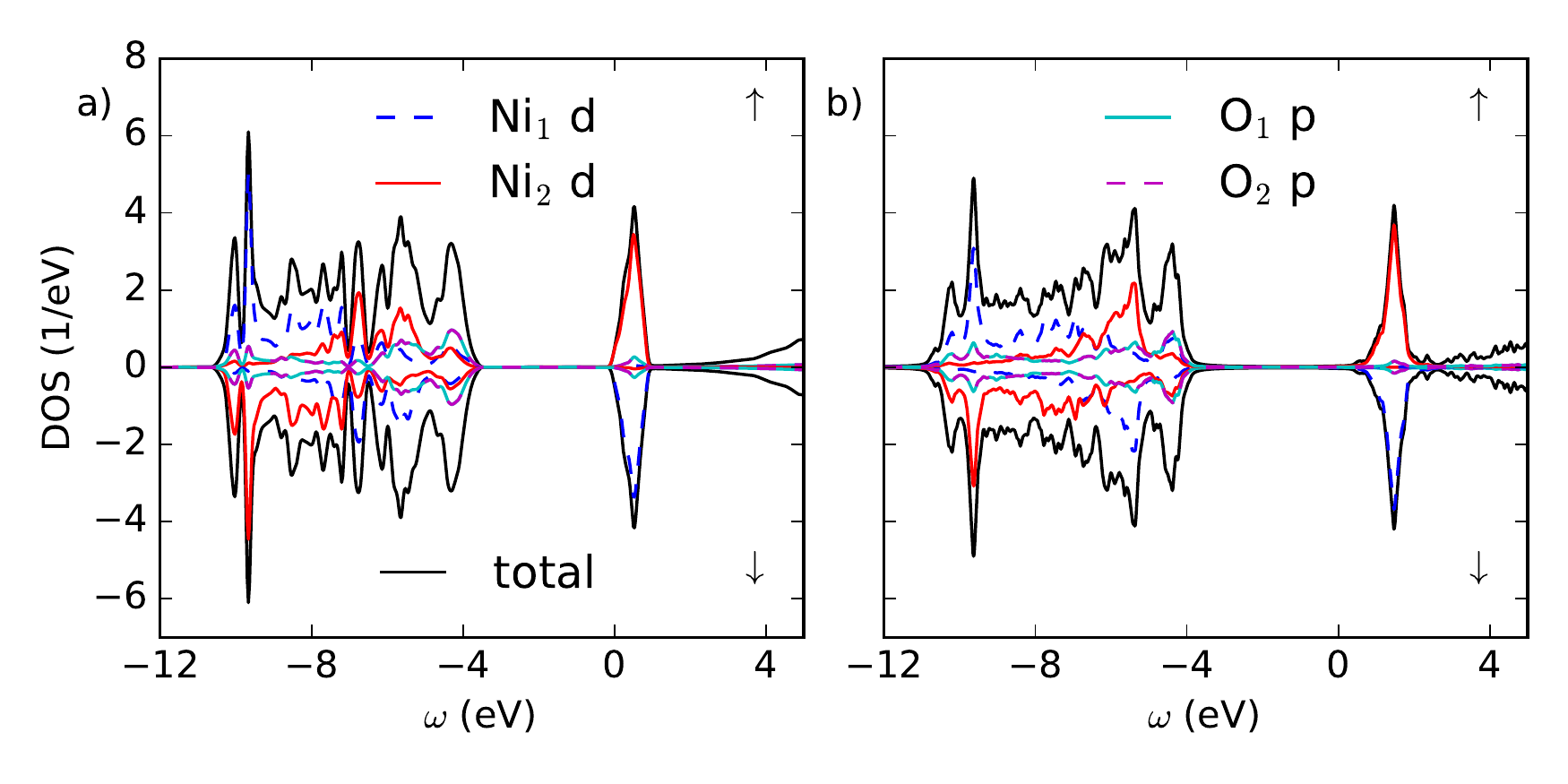}
\caption{Orbitally decomposed local density of states for antiferromagnetic NiO in a) LDA+U as implemented in VASP and b) in LDA++ using a charge self-consistent interface between VASP and an external Hartree-Fock solver.}
\label{fig:ldau_ldaext}
\end{figure}

With the new density $n(r)$ one can recalculate the DFT potential and solve the resulting Kohn-Sham Hamiltonian, which then defines a new Kohn-Sham Green function via Eq. \ref{eq:DefKSGreenfunction}. In this way, a charge self-consistent DFT++ scheme is obtained, which includes interactions between electrons of the correlated subspace and the rest in a fully self-consistent static mean-field manner. Several implementations of charge self-consistent DFT+DMFT have been reported, e.g. Refs. \cite{PhysRevB.84.054529,PhysRevB.86.155121,0953-8984-24-7-075604}, based on projector formalisms similar to Sec. \ref{sec:projector}. It is intuitively clear that the Hartree terms occurring within DFT++ charge self-consistency counteract large charge redistributions. In other words, ambiguities stemming for instance from the unknown double-counting potential can be expected to be less severe in charge self-consistent DFT++ calculations as compared to one-shot calculations. This has been explicitly demonstrated, e.g., 
for the Matsubara self-energies in the iron pnictide superconductor LaFeAsO, where the discrepancy between FLL and AMF approaches is significantly reduced in the fully charge self-consistent scheme \cite{PhysRevB.84.054529}. In nanosystems, the issue of charge redistributions can be even more severe than in bulk systems, since charge redistributions over larger distances are facilitated by the large supercells needed to describe the nanosystems. We thus implemented a charge self-consistent version of DFT+DMFT in the framework of the VASP code in collaboration with project P8. A simple benchmark is presented in Fig. \ref{fig:ldau_ldaext}, where we compare the density of states of antiferromagnetic NiO calculated from LDA+U (panel a)) and from a charge self-consistent LDA++ calculation where we solve the corresponding Hubbard model in mean field approximation. In principle, both methods should give the same density of states. Qualitatively, the orbital resolved DOS are the same. However, due to minor 
discrepancies in the definition of the projectors which are used internally in VASP and externally for the interface, some differences such as the gap and details in the valence band occur.

\section{Impurity solvers}
\label{sec:impsolv}

The solution of the Anderson impurity model for general parameters has to be done numerically by means of, e.g., quantum Monte Carlo \cite{CTQMC_RMP} (QMC), numerical renormalization group \cite{bulla_numerical_2007} (NRG), or exact diagonalization (ED) methods \cite{georgesRMP68}. While NRG and QMC are in principle numerically exact methods, they become computationally very demanding, when dealing with many orbitals, hybridization functions with low symmetry, spin orbit coupling and general fermionic four operator Coulomb vertices. Here we briefly describe the QMC and ED solvers that we mostly used for the study of nanoscopic systems.

\subsection{Continuous-Time Quantum Monte Carlo}
\label{sec:ctqmc}

The quantum Monte Carlo results presented here were obtained using the numerically exact continuous-time version of the algorithm using the hybridization expansion \cite{CTQMC_RMP}. We used the w2dynamics code package, which is a hybrid Python/Fortran90 code written in Vienna and W\"urzburg. It is a versatile code that is capable of handling complex systems with multiple impurities and difficult nano geometries. Impurities can be completely inequivalent and independently defined with their own Coulomb interactions and hybridizations. A general interface to any electronic structure code is provided via the usage of the Wannier Hamiltonian in reciprocal space as input.
The solver portion of the code is a highly optimized Fortran90 code with Segment \cite{segment_2006}, Matrix \cite{matrix_2006} and Krylov \cite{krylov_2009} implementations of the hybridization expansion algorithm, that are optimal in different regions of the parameter space. A general four index Coulomb tensor (e.g obtained from cRPA or cDFT) can be used as well. The code is capable of identifying certain conserved quantities \cite{parragh_2012} depending on the form of the interaction used (density-density, Kanamori, full) which greatly speeds up the simulations.

\subsection{Variational exact diagonalization approach}
\label{sec:varED}

The general concept of the exact diagonalization approach is to exactly solve an auxiliary Hamiltonian, which is an approximation to the full Hamiltonian in some systematic sense and is solved exactly. As solving a many-body problem exactly means diagonalizing the Hamiltonian on its Fock-space, which grows exponentially with the dimension of the Hilbert space, the original model has to be discretized in some way. Loosing the ``exactness'' of QMC has however some important advantages: first of all, ED has no restriction on the low symmetries and the general form of Coulomb vertices and, second, it allows us to access real-axis Green's functions and self-energies without the need of stochastic analytic continuation of noisy QMC data. 

Several efforts are dedicated to alleviate the problem of the bath discretization of ED \cite{zgid_truncated_2012,kananenka_systematically_2014,PhysRevB.90.085102,bauernfeind_fork_2016}.
Here, we briefly discuss a strictly variational method of approximating an AIM with continuous bath by an AIM with finite strongly reduced number of bath sites, which we call variational ED method \cite{PhysRevB.91.235142}. It guarantees an optimal approximation to the AIM for a given number of bath sites in the sense of thermodynamic ground-state properties. The method is based on the well-known Peierls-Feynman-Bogoliubov variational principle \cite{peierls,bogoliubov,feynman}. The minimization of the Peierls-Feynman-Bogoliubov free energy functional 
\begin{align}
\tilde{\Phi}[\rho_{\tilde H}] = \Phi_{\tilde H} + \langle{H-\tilde H}\rangle_{\tilde H},\label{eq:EDstar_functional}
\end{align}
leads to an optimal approximation of a Hamiltonian $H$ by an effective simpler model $\tilde H$ in terms of an optimal density matrix. Here $\Phi_{\tilde H} = -\frac{1}{\beta} \ln Z_{\tilde H}$ is the free energy of the effective system. $\langle \cdot  \rangle_{\tilde H}$ denotes a thermodynamic expectation value with respect to the effective system. In our case $H$ represents the full Anderson impurity model and $\tilde H$ is the model with discretized bath.

\begin{figure}\sidecaption
\includegraphics[width=0.65\columnwidth]{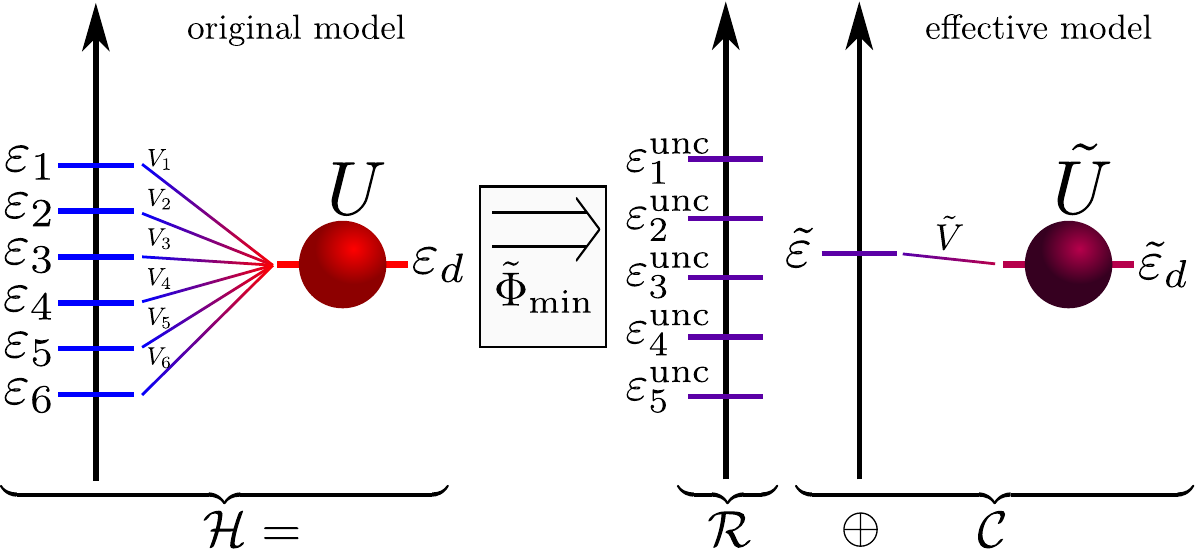}
\caption{(Color online) Illustration of the original and effective model for the case of one orbital and six bath sites. Blue represents bath character and red impurity character: In the effective model bath and impurity states can be mixed. Figure adapted with permission from Ref. \cite{PhysRevB.91.235142}. Copyrighted by the American Physical Society.}
\label{fig:benchBath}
\end{figure}

The ansatz for the effective model determines the quality of the method and has to be chosen carefully. It has to be simple enough that we can solve it exactly but still include explicit interaction terms. We divide the Hilbert space of the effective system into two classes: First, a small subset, the so-called correlated subspace $\mathcal{C}$, is equipped with interaction terms in the effective Hamiltonian $\tilde H ^\mathcal{C}$. The remaining space, the uncorrelated subspace $\mathcal{R}$, is equipped with single-particle terms only and described by $\tilde H ^\mathcal{R}$. The structure of the total effective Hamiltonian $\tilde H=\tilde H ^\mathcal{C}+\tilde H ^\mathcal{R}$ for the case of a single impurity orbital is depicted in the right panel of Fig. \ref{fig:benchBath}. In contrast to the original model (left panel of Fig. \ref{fig:benchBath}), the effective model consists of two decoupled parts: First, the effective interacting impurity coupled to one bath site only and second the remaining bath 
sites. For concreteness, we consider a cluster consisting of a multi-orbital impurity and one bath site per impurity orbital for the correlated space but other choices are similarly possible. The single particle states of the effective model are related to those of the original model by a unitary transformation $u_{ij}$, which allows for mixing of original ``bath'' and ``impurity'' character in the effective model. The optimal matrix elements of the effective model, as well as the optimal unitary transformation $u_{ij}$ are found by minimizing the functional (\ref{eq:EDstar_functional}).

The Hamiltonian $\tilde H ^\mathcal{C}$ states a many-body problem which can be solved by exact diagonalization. In contrast, the Hamiltonian
$\tilde H ^\mathcal{R}$ states a single-particle problem and can be solved trivially. In summary, the Hamiltonian $\tilde H = \tilde H^\mathcal{C} + \tilde H^\mathcal{R}$ defines an effective Hamiltonian, which can be solved exactly and thus $\tilde \Phi$ (Eq. (\ref{eq:EDstar_functional})) can be calculated exactly.

Finally, we note that the amount of variational degrees of freedom in the variational ED approach is such that it includes Hartree-Fock as the limiting case $\tilde  U_{\alpha\beta\gamma\delta} \to 0$. Thus, we expect that variational ED will generally give more accurate energy estimates than Hartree-Fock. Details on how to deal with the large number of variational degrees of freedom in the variational ED approach are discussed in Ref. \cite{PhysRevB.91.235142}.


\begin{figure}[htb]\sidecaption
\includegraphics[width=0.5\columnwidth]{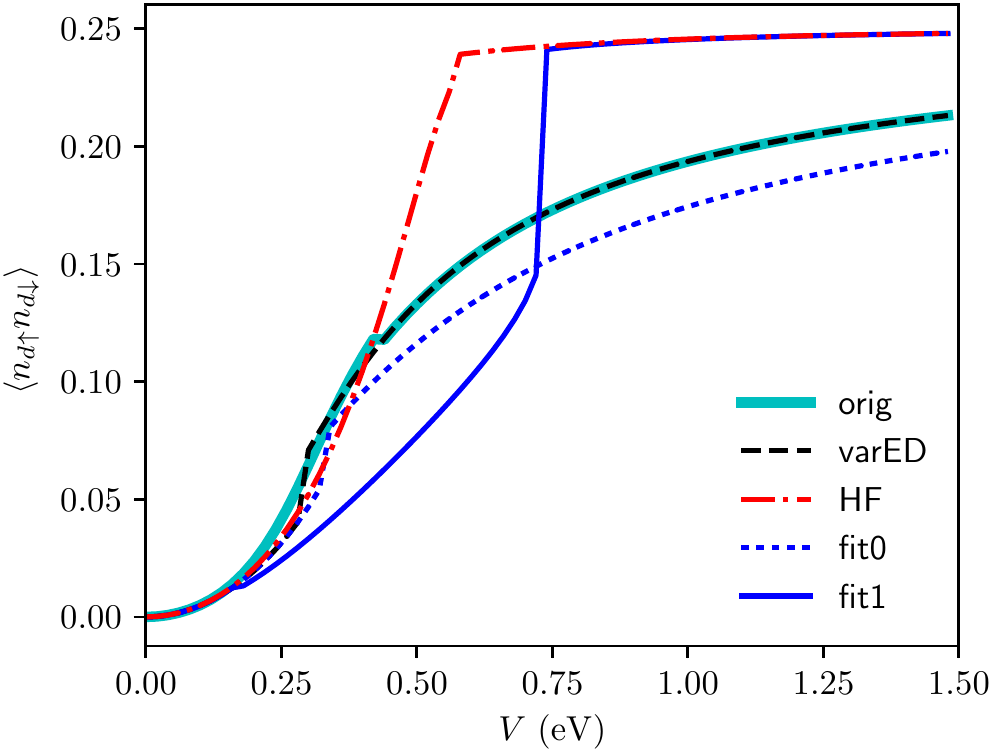}
\caption{(Color online) Benchmark against an exact solution for single orbital Anderson impurity models with hybridization matrix element $V$. Comparison of $\langle n_{d\msu} n_{d\msd} \rangle$ obtained from an exact solution (``orig'', bold cyan) and calculated by the four different approximate methods, i.e., the variational ED method (``varED'', solid black), unrestricted Hartree-Fock (``HF'', dashed red) and fits of hybridization functions on the imaginary axis with different weight functions ($W_n=1$: ``fit0'', dotted blue; $W_n=1/\omega_n$: ``fit1'', dashed blue.)}
\label{fig:benchmarkED}
\end{figure}
We demonstrate the performance of the variational ED method by applying it to a system (schematically shown in the left panel of Fig. \ref{fig:benchBath}) with only six bath sites which is small enough to be solved exactly. In Fig. \ref{fig:benchmarkED} we compare the double occupation obtained from the variational ED, Hartree-Fock and the fit of Green functions on Matsubara frequencies \cite{caffarel_exact_1994} with two different weighting functions ($W_n=1$, ``fit0'' and $W_n=1/\omega_n$, ``fit1'', compare Ref. \cite{senechal_bath_2010}). The parameters of the model are the following: The impurity level is $\varepsilon_d=-2.0\eV$, the interaction strength is $U=4.0\eV$. The six bath levels are equally aligned around a mean bath energy $\varepsilon_b = 0.02 \eV$ in an interval of $2\eV$ (i.e. the bandwidth of the bath). The hybridization is swept from $V_k=0.0\eV$ to $V_k=1.5\eV$. Small hybridization ($V=0$) contains the atomic limit, large hybridization ($V\rightarrow\infty$) is the non-interacting limit. The region in between ($V\sim 
0.3 \eV $) is strongly correlated as interaction and kinetic energy are on the same order. The Hartree-Fock method gives satisfactory results for small hybridization strengths and reproduces the non-interacting limit. The choice of the weight function in the fit methods has a clear impact on how well the double occupancies are reproduced, which reveals the ambiguities associated with the "conventional" bath discretization schemes "fit0" and "fit1". The variational ED method, in contrast, gives an unambiguous and very accurate approximation to quantities like double occupancies and outperforms all other methods. Small deviations to the exact double occupation exist in the strongly correlated region around $V=0.3\eV$. Including explicit interaction terms in the effective Hamiltonian leads to a drastically better performance than Hartree-Fock. The same conclusion also holds for estimates of total energies, which makes the variational ED approach very interesting in the context of materials simulations.

In Ref. \cite{PhysRevB.91.235142} it is shown that the method is indeed of practical use for realistic multi-orbital impurities. There, an ab-initio derived model of Co in Cu is solved and compared to a quantum Monte Carlo solution.

\section{Applications}
\label{sec:applications}

We now review applications of the compuational scheme outlined in the two previous sections. We start with a general discussion electron correlations, charge fluctuations and magnetism in multi-orbital Anderson models and consider two examples of nanosystems without any translational symmetry: so-called "Hund's impurities" as realized by hydrogenated Fe on Pt(111) (section \ref{subsec:Hund}) and metal organic molecules on surfaces as complex Kondo systems (section \ref{subsec:molecules}). Finally, we discuss the electronic structure of the Cr (001) surface in section \ref{subsec:Cr001}, as an example of a correlated system which retains translation symmetry in two dimensions.

\subsection{Electronic correlations driven by Hund's exchange interaction}
\label{subsec:Hund}



There is a wide class of transition metal compounds in which Hund's rule coupling leads to exotic electronic properties. Examples of these so-called ``Hund's metals'' \cite{Hund_Georges_Review_2013} include unconventional superconductors such as iron pnictides and chalcogenides \cite{Stewart:2011,Yin:2011,Werner2012} as well as non-Fermi liquids as realized in ruthenates \cite{Werner2008,Mravlje2011,deMedici2011}. Within the reasoning of DMFT, magnetic impurities strongly coupled to the electronic states of a metallic host can be viewed as fundamental constituents of such Hund's metals and are called ``Hund's impurities'' \cite{huang_electronic_2014,HundImp_2015}.


This idea is described in the following for the particular case of a 3\textit{d} transition metal adatom on a metallic substrate following Refs. \cite{huang_electronic_2014,HundImp_2015}. If the adatom is still in the gas phase an integer number of electrons is filled into the five 3\textit{d} orbitals according to Hund's first rule: The orbitals are first filled up by electrons having the same spin, before being filled with the remaining electrons of opposite spin. This is driven by the intra-atomic exchange energy, or so-called Hund's rule exchange $J$. After adsorption on the metallic substrate, electrons can hop on or off the adatom's orbitals into the bath of substrate conduction electrons, which has an electronic density of states $\rho$, paying or gaining the direct on-site Coulomb energy $U$. This hopping leads to charge fluctuations at the impurity site. The hopping term $V$~\cite{Carbone2010} and the valency~\cite{Gardonio2013} determine whether the adatom is most appropriately described in terms 
of atomic multiplets, itinerant electrons, or a mixture of both with distinct correlation effects. 

\begin{figure}%
\includegraphics[width=\columnwidth]{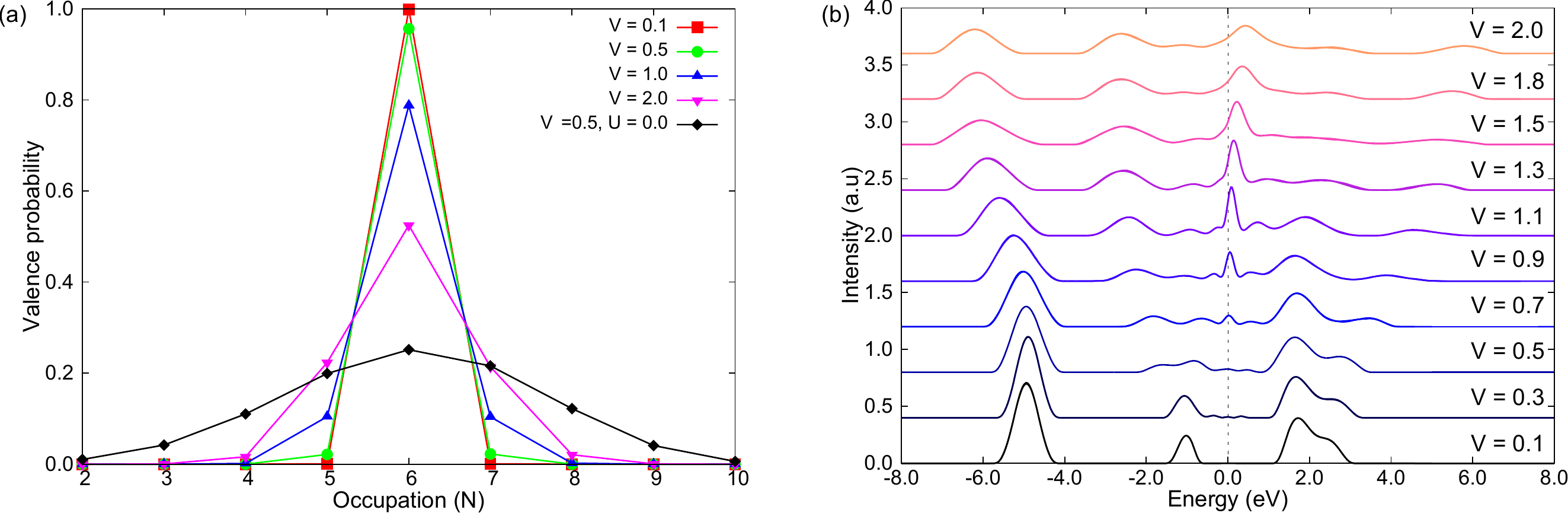}%
\caption{Valency (a) and electronic excitation spectra (b) of a five orbital Anderson impurity with local interactions $U=4$\,eV and $J=1$\,eV at filling $N=6$ as calculated by CT-QMC. Figure adapted with permission from Ref. \cite{huang_electronic_2014}. Copyrighted by the American Physical Society.}%
\label{fig:HundImp}%
\end{figure}

This is illustrated for an orbitally degenerate five orbital Anderson impurity with local interactions $U=4$\,eV and $J=1$\,eV and bath density of states $\rho=0.05$\,eV$^{-1}$ at filling $\langle \hat N\rangle=6$ in Fig. \ref{fig:HundImp}. For vanishing hybridization $V\approx 0.0$ the system has a well-defined valency and (half-) integer quantized spin. Also with weak hybridization, $V\ll 1$\,eV, the adatom essentially retains its integer valency. With hybridization increased beyond $V\gtrsim 1$\,eV sizable charge fluctuations set in, as the valence histogram (Fig. \ref{fig:HundImp}a) demonstrates. If the adatom had only a single orbital, the magnetic moment would be simply quenched in this case. However, in the multi-orbital case, there exists a particular regime with strong hybridization which is yet too weak to overcome the Hund's coupling $J$, where strong charge fluctuations can coexist with sizable local magnetic moments. This has been termed the Hund's impurity regime~\cite{huang_electronic_2014,HundImp_2015}. The electronic excitation spectra (Fig. \ref{fig:HundImp}b) display a characteristic evolution from the atomic limit to the Hund's impurity regime~\cite{huang_electronic_2014}: When switching on the hybridization, a Kondo-type low energy resonance appears, broadens and
merges with some of the upper and lower Hubbard peaks. In contrast to the single-orbital case, several atomic multiplet peaks persist and additional exchange split
satellites appear in presence of strong charge fluctuations.

The effective charging energy $U_{\rm eff}$ of a multiorbital impurity system, i.e. the energy which determines the amount of charge fluctuations in the system, can be estimated by the formula $U_\text{eff}^{(N)}=E(N+1)+E(N-1)-2E(N)$, where $E(N)$ is the lowest atomic energy of the $N$ electron state. For rotationally invariant interaction, this gives $U_\text{eff}^{(N=5)}=U+4J$ (half-filling), $U_\text{eff}^{(N=6,9)}\approx U-3J/2$ and $U_\text{eff}^{(N=7,8)}\approx U-J/2$.~\cite{huang_electronic_2014} Hence, particularly Fe based impurity systems which typically have $N\approx 6$ are expected to realize strong charge fluctuations and Hund's impurity behavior. 

The case of hydrogenated Fe, FeH$_x$, on Pt (111) has been investigated in a collaboration between experiments and FOR 1346 as reported in Ref. \cite{HundImp_2015}. FeH$_x$ on Pt (111) realizes strong charge fluctuations coexisting with local magnetic moments for several binding geometries and different degrees of hydrogenation, $x=0,1,2$. This leads to non Fermi liquid electronic self-energies due to spin scattering at intermediate temperatures, as shown in \cite{HundImp_2015}. While this intermediate temperature non Fermi liquid behavior is very generic for Hund's impurity systems, FeH$_x$ on Pt (111) presents a model system where the low temperature magnetic properties can be strongly controlled by changing the binding site and degree of hydrogenation of the Fe adatoms with the tip of a scanning tunneling microscope. In this way, it turned out to be possible to control four almost degenerate energy scales (Zeeman, thermal, Kondo and magnetic anisotropy energy) and to tune the Hund's impurities from 
realizing an emergent unquenched magnetic moment to a multi-orbital Kondo state. The Kondo state is particularly interesting, here, as it emerges without well quantized (half-) integer local moments at the Fe sites. Hence, Ref. \cite{HundImp_2015} shows that Kondo screening is possible without quantization
of the local moments. The coupling of electrons to fluctuating magnetic moments is regarded a possible mechanism for superconductivity at high temperatures. Therefore, (arrays of) Hund's impurities could provide a bottom-up way for the understanding of the complex physics of systems like iron pnictide and chalcogenide superconductors.

\subsection{Transition-Metal Phthalocyanine on Ag(001)}
\label{subsec:molecules}

\begin{figure}
\resizebox{0.99\columnwidth}{!}{%
  \includegraphics{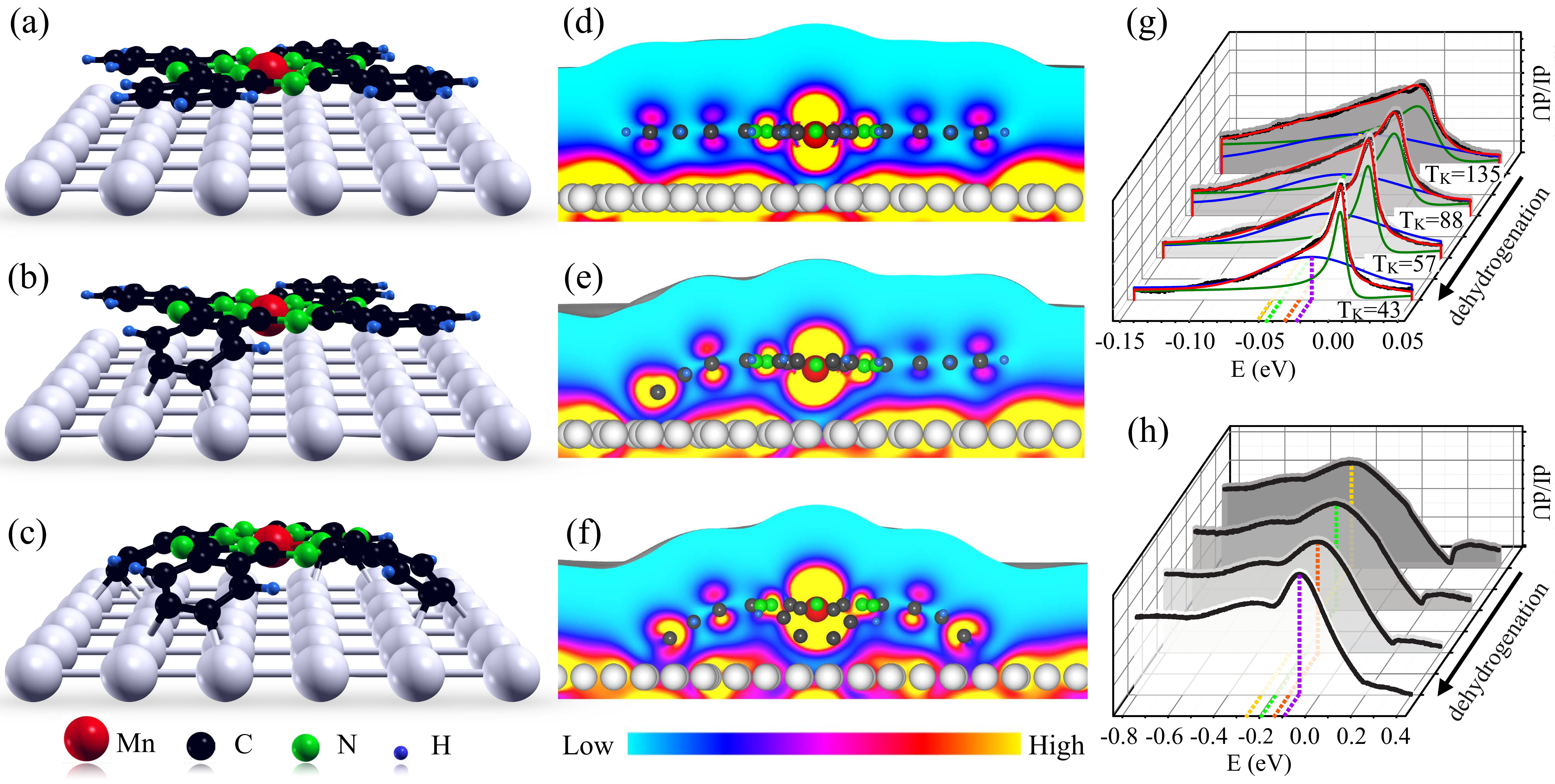} }
\caption{(a-c) Fully relaxed structures of MnPc and dehydrogenated MnPc(-2H) and MnPc(-8H) on Ag(001). (d-f) Corresponding partial charge densities obtained by integrating the band structure in a window of $\pm 200meV$ around the Fermi level. (g-h) Series of tunneling spectra recorded for the complete and increasingly dehydrogenated MnPc and FePc molecules respectively. Adapted from Ref. \cite{state_prb_2015}.}
\label{fig:mnpcfig2}       
\end{figure}

In this section we focus on transition-metal phthalocyanine molecules (TMPc) on the (001) surface of silver.
This system is an example of how electronic and transport properties of atomic and molecular conductors can be controlled by manipulation of the molecular geometry \cite{iancu,choi}. It is conceivable to use similar molecular magnets\cite{gatteschi} as building blocks in the quest for improved magnetic storage devices and/or for spintronics applications \cite{wolf2001,zutic2004,bogani2008,schmaus2011}.
Metal–organic molecules such as transition-metal Pc are particularly well suited for theoretical studies of strong electronic correlations, since they contain a single transition metal atom at the center and, additionally, handle very well in experiments. When deposited on the Ag (001) surface, see Fig. \ref{fig:mnpcfig2}(a) for a generic structure, certain representatives, NiPc and CuPc, exhibit a Kondo effect which appears to be localized on the arms of the organic ligand of the transition metal ion \cite{mugarza_2012}.
Instead, MnPc exhibits the Kondo effect in the $3d$ shell of Mn \cite{Kuegel_MnPc}. Additionally, it is possible to manipulate the structure of the ligands with an STM tip \cite{state_prb_2015}, thus gaining some degree of control over the physics.

When faced with such a challenging problem as the calculation of electronic correlation effects in MnPC on a surface two approaches come to mind. One would be the quantum chemical approach to attack the full system including all interactions using a truncated configuration interaction or similar approach to account for the correlation effects \cite{helgaker}. The other one, more popular among physicists, is to "truncate" the problem already at the level of the Hamiltonian and then solve the remainder as exactly as possible. We will follow latter route here, observing that the system can be viewed as a realization of an Anderson impurity model \cite{anderson_localized_1961}, if one assumes that everything surrounding the correlated central atom can be reasonably described within a one particle theory, such as density-functional theory (DFT). The model was originally designed and is best suited for the description of $d$ or $f$ shell impurities in simple metal hosts and is capable of describing local 
correlation physics, 
like the Kondo effect \cite{Schrieffer_Wolff,Hewson}. Since the remainder of the system consists of a simple metal with $sp$ electrons close to the Fermi level (Ag) and reasonably uncorrelated elements (C, N and H) such an approach can be justified. They comprise what is called the dynamical, i.e., energy dependent, hybridization, whereas the impurity levels or crystal field constitute the static part. In our case the \textit{bath} does not only consist of the substrate, but also involves the organic parts of the molecule. 

\begin{figure}
\resizebox{0.99\columnwidth}{!}{%
  \includegraphics{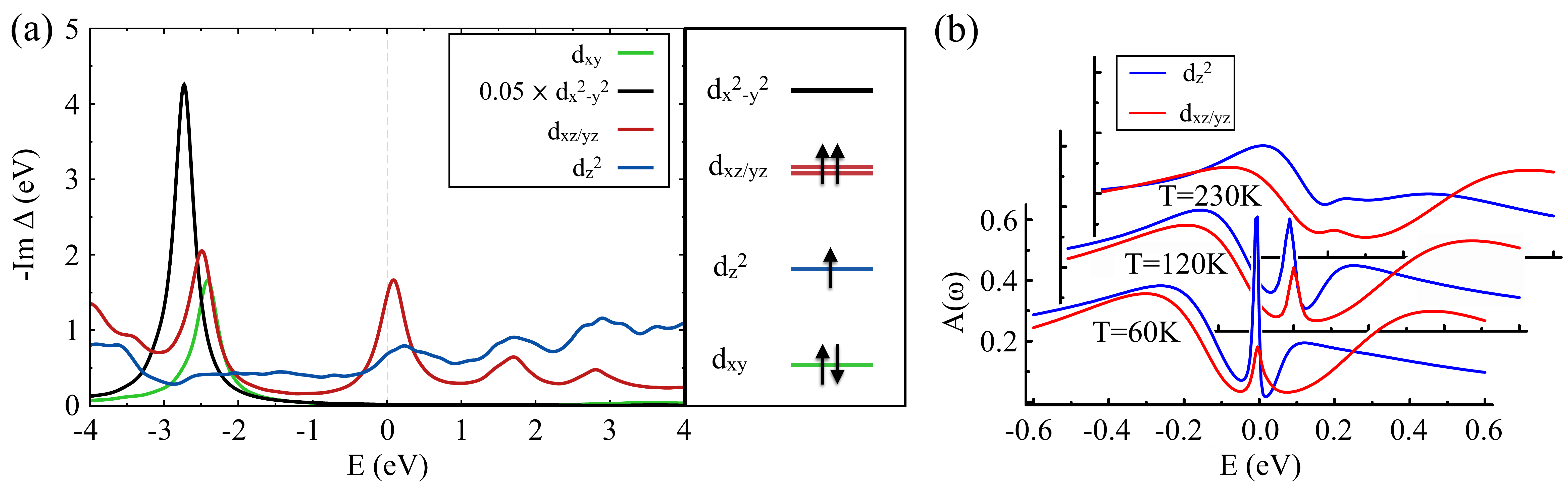} }
\caption{(a) Imaginary part of the hybridization function along with its (schematic) static real part (i.e. the crystal field splitting). The dominant configuration of the Mn atom, i.e. $d^5_{S=3/2}$ is indicated. (b) Temperature dependence of the DFT++ spectral function of part of the Mn $3d$ shell. Reprinted with permission from Ref. \cite{Kuegel_MnPc}. Copyright 2014 American Chemical Society.}
\label{fig:mnpcfig1}       
\end{figure}

In our treatment the hybridization is given by a continuous orbital and energy dependent function, $\Gamma_{ij}(\omega)=\sum_{\nu}\frac{ V^{\phantom{\ast}}_{\nu i} V_{\nu j}^\ast}{\omega-\varepsilon_{\nu}+i\delta}$, which can be obtained from DFT as described in section \ref{sec:interface}.
The final ingredient in the calculation, after the local basis and the hybridization function, is the local Coulomb interaction. The Coulomb interaction is in general a rotationally invariant fourth-rank tensor \cite{coury_2016} and there is no mystery about its general form and entries for atoms \cite{SlaterBook}. It is, however, not so simple to calculate the entries of the tensor for a solid or a nanodevice, due to dynamical screening \cite{aryasetiawan_2006}. We have used the density-density portion of the full Coulomb interaction as obtained from the Slater integrals $F^0, F^2$ and $F^4$ for $3d$ shell systems \cite{SlaterBook}. The numerical values we used were $F^0 = 4.5$eV, $F^2=6.03$eV and $F^4=3.77$eV which were obtained for MnPc by means of the constrained random-phase approximation (cRPA) \cite{jacob_2013}. To solve the model we employed the CT-QMC solver as described in section QMC.
Single MnPc molecules on Ag(001) nicely illustrate the general features of a correlated molecular system in contact with metallic leads or a metal surface. The structure of a MnPc molecule on the Ag(001) surface is shown in Fig.~\ref{fig:mnpcfig2}a. Before any Anderson model calculations one has to establish the structure of the molecule and the adsorption site as well as possible changes in the geometry. A reliable approach is to perform a relaxation within DFT employing a GGA functional amended by a dispersion correction to account for van-der-Waals interactions. From experience a good combination is PBE+D3, which corrects the under-binding tendencies of pure PBE \cite{grimme_2011}. LDA can, due to miraculous error cancellations, also perform quite well in such a case, see e.g. \cite{rohlfing_2008}. 
When the structure is established and a localized basis is available either directly or via a Wannier construction the hybridization function and static crystal field can be calculated. Both are shown in Fig.~\ref{fig:mnpcfig1}a. The hybridization function is very different for each $d$ orbital and exhibits the generic features of a molecular system in contact with a surface or leads. One can make out three different types of $d$ orbitals: 

\begin{itemize} 
\item orbitals that couple only to the molecular states, like the $d_{xy}$ and $d_{x^2-y^2}$. Their hybridization function shows strong peaks at specific energies and closely resembles the hybridization function calculated for the free molecule, see Ref. \cite{bhandary_2016}.

\item orbitals that couple strongly to the surface and only weakly to the molecular states, like the $d_{z^2}$. Its hybridization is mostly smooth and almost constant as expected from the "flat" $sp$-electron density of states of the silver substrate.

\item orbitals that couple moderately to both the molecular states as well as the surface, like the $d_{xz/yz}$.
\end{itemize}

Due to their symmetry and orientation in space such a division of the orbitals is typical for a metal–organic molecular system in contact with metallic leads or surfaces. Due to the square planar structure of the molecule, as well as the presence of additional Ag atoms underneath, the level crystal field splitting of the $3d$ shell is as shown in Fig.~\ref{fig:mnpcfig1}. The $d_{z^2}$ orbital, as well as the $d_{xz/yz}$ orbitals are half-filled and are as such prime candidates for correlated electron behavior. The level positions are, however, different for different TMPc species, due to different adsorption distances. 
Having all the ingredients of the Anderson model now at hand we can solve the model and calculate observables like the spectral function or self-energy. Since these types of metal–organic molecules already exhibit a complex electronic structure without the surface \cite{bhandary_2016}, we expect some interesting many-body effects. Indeed below a certain temperature a sharp peak appears in the spectral function of the $d_{z^2}$ orbital, as shown in Fig.~\ref{fig:mnpcfig1}b. Corroborated by a Fermi-liquid behaviour of the electronic self-energy, we can identify this as the signature of the Kondo effect. Such a spectral feature is also observed in STM experiments \cite{Kuegel_MnPc}. For the $d_{xz/yz}$ orbitals a similar feature appears, which can be shown to be of non-Fermi liquid character, however. In FePc and CoPc, one would also expect such an effect due to the non-vanishing spin on the $3d$ shell, however, in experiments no low energy spectral feature indicative of a Kondo effect is found. This can be 
explained by the 
increasing filling of the $d_{z^2}$ orbital, which is effectively full in Fe and Co, leaving the local moment regime of the Anderson model \cite{Kuegel_MnPc}.
This is already interesting on its own, but MnPc allows us (and experimentalists!) to do more. The Kondo effect in MnPc can to some extent be controlled by manipulating the structure of the molecule by a process called dehydrogenation \cite{state_prb_2015}. As the name suggests hydrogen atoms are removed using an STM tip from the molecule employing a voltage pulse \cite{zhao2005}. The effect on the structure can be simulated within DFT, the result of removing two and eight hydrogen atoms are shown in Fig.~\ref{fig:mnpcfig2}b and c. The concomitant changes in the electronic density are shown next to the respective structures in Fig.~\ref{fig:mnpcfig2}e and f. The downward bending of the arms of the molecule is accompanied by a depletion in charge density close to the newly formed bonds as also observed in experiment \cite{state_prb_2015}. The bending also leads to an increase in distance between the central atom and the surface, which in turn changes the local electronic structure in multiple ways. It reduces 
the hybridization between central 
atom and substrate, especially for the 
$d_{z^2}$ orbital, and also leads to a uniform shift to smaller binding energies of the partially filled $d_{xz/yz}$ and $d_{z^2}$ orbitals. The former leads to a decrease in the Kondo temperature as shown in Fig.~\ref{fig:mnpcfig2}g, while the latter can be seen in experiment as an upward shift of the states superimposed over the Kondo resonance. Both effects occur also in FePC, see Fig.~\ref{fig:mnpcfig2}h, the reduction in coupling to the substrate manifests as a general sharpening of spectral features, since a Kondo effect is absent here.

We have discussed the Kondo effect in TMPc molecules on Ag(001) as well as the effects of dehydrogenation seen in STM based mainly on local quantities of the molecules calculated within the respective Anderson impurity models. In the systems presented here this turned out to be sufficient for the qualitative description of the physics. However, in an STM experiment the tip and the resulting interaction between tip and sample can play a crucial role as well. In particular, regarding the understanding of transport through a correlated molecule measured by STM, it was realized in Ref. \cite{tpp-paper}, that not only the local correlations on the molecule are important, but also the voltage drop over the junction is critical in understanding the line-shapes in the conductance.

\subsection{Electronic structure of the Cr(001) surface}
\label{subsec:Cr001}
Bulk Cr crystallizes in a body-centered crystal (bcc) structure \cite{ross_high_1963}.
Neutron diffraction experiments \cite{shirane_neutron_1962,arrott_neutron-diffraction_1967} reveal a spin-density wave ordering of bulk Cr with a magnetic moment of about $0.6\mu_B$ per atom at low temperatures and a N\'eel temperature of approximately $T_N^\text{bulk}\sim311\ \text{K}$ \cite{rapp_electrical_1978}.

\begin{figure}[htb]\sidecaption
\centering
\mbox{
\includegraphics[width=0.55\columnwidth]{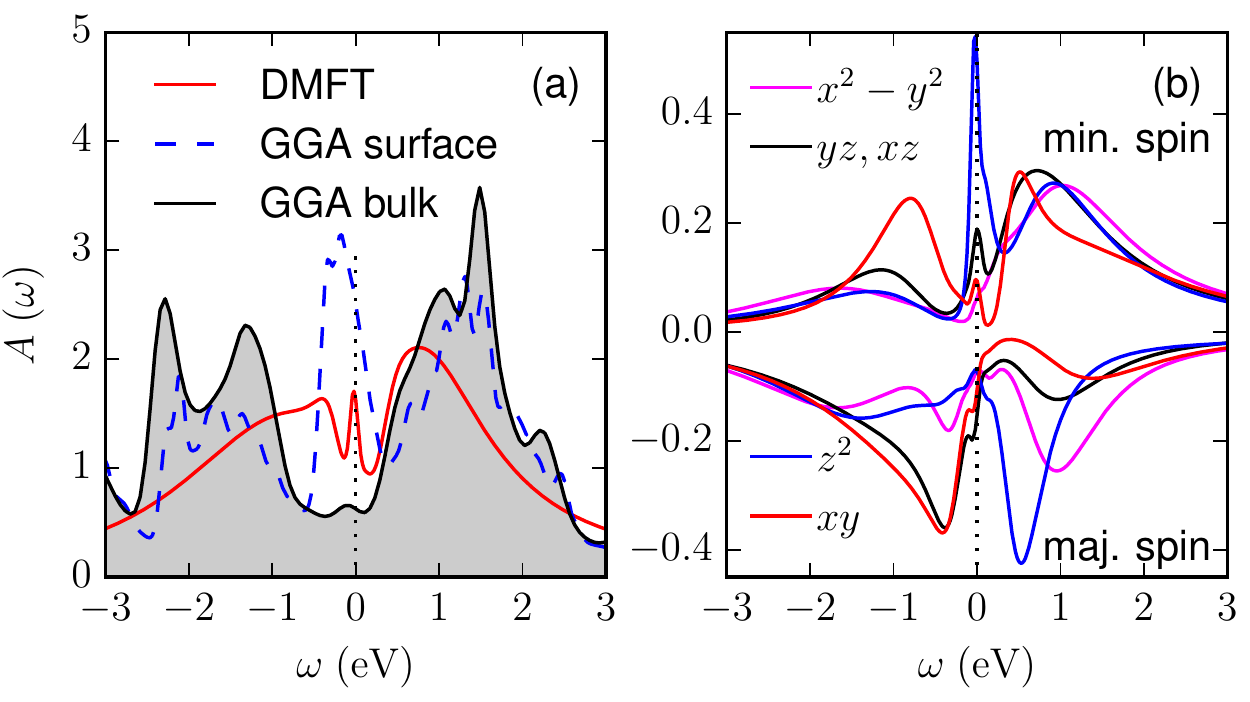}
}
\caption{(Color online) (a) Orbital and spin averaged local density of states (LDOS) from LDA+DMFT simulations at the temperature $\beta=40\eV^{-1}$ (red), and GGA (dashed blue). The GGA DOS of bulk Cr is shown in black. (b) orbital and spin resolved LDOS of the surface atom from LDA+DMFT. Figure adapted with permission from Ref. \cite{schuler_many-body_2016}. Copyrighted by the American Physical Society.}
\label{fig:totalDOS}
\end{figure}

Considering the lattice of antiferromagnetic Cr it is obvious that a cut along the (001) direction leads to a surface where atoms of layers parallel to the surface are ferromagnetically aligned. Thus, the first layer of the (001) surface consists of ferromagnetically aligned atoms. Intriguingly, the surface magnetic state persists up to much larger temperatures ($T_N^\text{surf} \sim 750-800 \ \text{K}$) than the bulk state, as measured in Refs. \cite{klebanoff_experimental_1985,klebanoff_observation_1984} by angular resolved photo emission (ARPES) and Ref. \cite{matsuo_magnetic_1980} by magnetization measurements with Cr particles of different diameter. This large magnetic moment originates from the altered paramagnetic electronic structure: The surface introduces massively more states close to the Fermi energy (compare Fig. \ref{fig:totalDOS} for a comparison of the bulk DOS and the surface LDOS), which spin split due to the exchange interaction and lead to a large surface moment \cite{leroy_electronic_2013}. 

\textit{Ab initio} based constrained random-phase approximation (cRPA \cite{cRPA_2004}) calculations for the Cr(001) surface \cite{csacsiouglu_strength_2012} show that the Cr $d$ electrons are subject to rather strong Coulomb interactions. This, together with the increased density of states at the surface lays the ground for correlation effects.

The question if correlations actually are important for the electronic structure has been subject of measurements using (inverse) photo emission spectroscopy (PES) (\cite{klebanoff_observation_1984,klebanoff_investigation_1985,klebanoff_experimental_1985,adhikary_complex_2012}) and scanning tunneling spectroscopy (STS) \cite{kolesnychenko_surface_2005}. These measurements all reveal a sharp peak close the Fermi energy. The origin of the peak was so far discussed either to be of single particle nature (surface state, \cite{budke_surface_2008}) or of many-body nature (orbital Kondo effect \cite{kolesnychenko_surface_2005}). Here, we review our work from Ref. \cite{schuler_many-body_2016} where we use the LDA+DMFT method to unravel the nature of the resonance in the electronic spectrum of Cr(001).  We show that local electronic correlations play a key role for understanding the electronic structure of Cr(001).

In order to reproduce the structure of the spectrum realistically, we model the Cr(001) surface by a slab of ten Cr atoms, which was partially relaxed using GGA calculations, and employ the projector formalism of LDA+DMFT described in Sec. \ref{sec:projector}. We solve the resulting multi-orbital Hubbard model of the slab of ten atoms by multi-site DMFT which allows for spatially inhomogeneous Coulomb interaction and antiferromagnetic ordering \cite{georgesRMP68}. Compare, e.g., Refs. \cite{lechermann_towards_2015} and \cite{valliPRL104} for a similar approach. 

To solve the impurity problems we use the CTQMC algorithm presented in Sec.~\ref{sec:ctqmc}. We cope with the double-counting problem by the requirement that the total occupation on each impurity obtained from the DMFT Green function $G$ matches the corresponding occupation obtained from the non-interacting bath Green function. This is called trace double-counting correction and reads
$ \Tr \rho_{\alpha\beta}^\text{imp} \stackrel{!}{=} \Tr \rho_{\alpha\beta}^{0,\text{loc}} $,
where $\rho$ is the density matrix for each atom. This scheme allows for a flexible way to change the double counting, namely to enforce a desired electron number on the Cr atoms. Here, we have enforced 4.5 electrons on each Cr atom.


The local density of states at $\beta=40\eV^{-1}$ summed over the $d$ orbitals and spins for the surface atom is shown in comparison to the GGA spectra in Fig. \ref{fig:totalDOS} (a). For the surface atom we can observe a three-peak structure resembling the experimental situation (compare Ref. \cite{budke_surface_2008}).

The orbital characters of the spectrum are presented in Fig. \ref{fig:totalDOS} (b). The resonance is dominated by a feature in the minority spin channel with $d_{z^2}$ character but also carries some spectral weight from the other $d$ orbitals, particularly the $d_{xz}$ and $d_{yz}$ orbitals. For $\beta=40\eV^{-1}$ the $d_{z^2}$ orbital also shows a sharp resonance in the majority spin channel. Overall, the orbital contributions of the spectra are in line with experimental findings in Ref. \cite{budke_surface_2008}.

To answer the question of how important correlation effects are at the surface and in the bulk, we calculate the quasiparticle weight $Z = \left[1 - \left.{\partial\real \Sigma(\omega)}/{\partial \omega}\right|_{\omega=0}\right]^{-1}$ for all orbitals and spins. Analyzing $Z$ for all layers of the slab, as done in \cite{schuler_theoretical_2016} shows that correlation effects are generally strongest at the surface. Also, the $d_{yz,xz}$ orbitals show the smallest quasiparticle weight, i.e., the strongest many-body renormalizations.

The nature of the resonance seems to be the combination of the rearrangement of the electronic structure at the Cr(001) surface (massively more states at the Fermi energy in contrast to bulk Cr) and the multi-orbital interaction effects, especially in the $d_{z^2}$ and $d_{xz/yz}$ orbitals. The interaction introduces two major effects: first, the spin dependent splitting of mainly the $d_{xz/yz}$ orbitals and, secondly, the appearance of quasiparticle peaks in $d_{z^2}$ and $d_{xz/yz}$ orbitals. We conclude that the resonance is a complex many-body effect in the $d_{z^2}$ and $d_{xz/yz}$ orbitals due to dynamic local-correlation effects which are directly related to the (single-particle) electronic structure at the surface.

\section{Non-local interactions and correlations}
\label{sec:non_loc}

\subsection{Realistic description of Coulomb interactions and screening in complex nanosystems: The Wannier function continuum electrostatics approach}
\label{subsec:wannier_electro}


The interaction terms entering the LDA+DMFT modeling can be in principle also derived from ab-initio calculations. However, special care on the issue of screening has to be taken here since the correlated subspaces involve some low energy sector of bands only. Hence, realistic interaction matrix elements entering the modeling should be appropriately screened. I.e. they should account for screening due to those states which are not explicitly treated in the low energy models. Depending on the later many-body treatment one has to ensure that screening channels are neither omitted nor doubly counted in the end. To calculate realistic appropriately screened Coulomb interaction matrix elements from first principles the so-called constrained Random Phase Approximation (cRPA) has been put forward.~\cite{cRPA_2004} It relies on excluding certain screening channels explicitly in a many-body perturbation theory calculation and its computational demand is comparable to GW calculations. Hence, direct cRPA calculations of complex nanostructures or also two-dimensional (2d) materials on substrates with large moir{\'e} supercells are often computationally unfeasible. To solve this problem, we developed an approach called ``Wannier function continuum electrostatics'' (WFCE)~\cite{WFCE_2015} which will be explained in the following with the example of layered materials and heterostructures.

The essential idea of WFCE is to combine ab-initio cRPA calculations for the bulk of a layered material, which are of moderate computational cost, with continuum
medium electrostatics to derive effective Coulomb interaction matrix elements in terms of Wannier functions for freestanding 2D materials or 2D materials embedded in complex dielectric environments. In this way, we can avoid supercell calculations involving complex environments or large vacuum volumes on the ab-initio side, which are numerically often very expensive. 

\begin{figure}%
\includegraphics[width=\columnwidth]{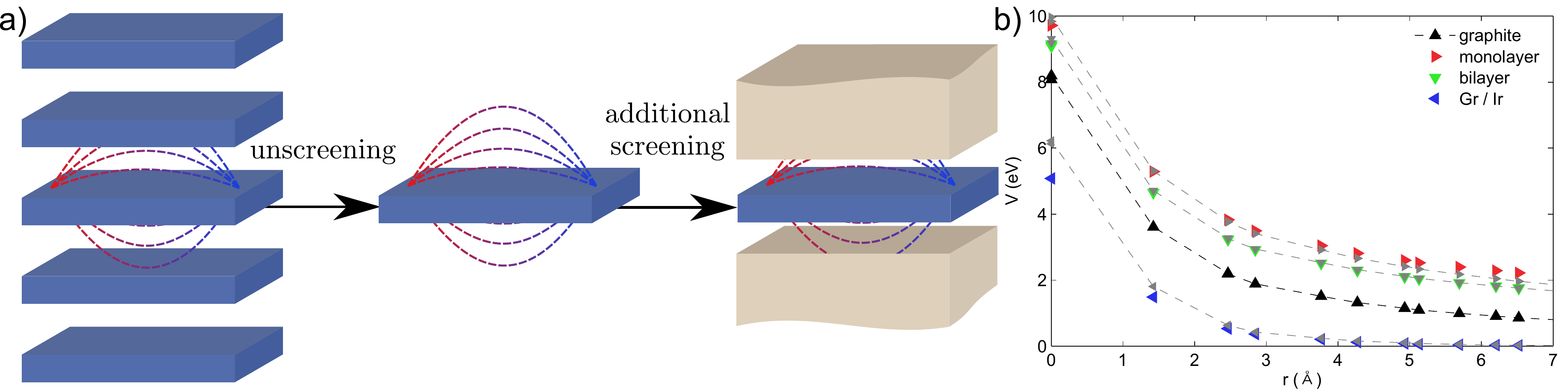}%
\caption{(a) Illustration of the WFCE idea. The modified screening when going from a bulk material to a monolayer or some more complex heterostructure is accounted for by means of macroscopic electrostatics. (b) Density-density matrix elements of the screened Coulomb interactions for graphite, monolayer and bilayer graphene as well as graphene embedded in Ir metal (Gr/Ir) in real space, i.e. as function of distance $r$. The screened interactions are generally retarded, i.e. frequency $\omega$-depedent. Here, the static limit $\omega=0$ is considered. Colored markers show ab-initio values, gray markers connected by dashed lines show the WFCE values. Panel b adapted with permission from Ref. \cite{WFCE_2015}. Copyrighted by the American Physical Society.}%
\label{fig:WFCEscheme}%
\end{figure}

Going from a bulk material to a monolayer or a complex heterostructure means changing the polarizability of the environment of the material of interest, as illustrated in Fig. \ref{fig:WFCEscheme}a). These changes have to be accounted for in the effective dielectric function of the system, $\mathbf{\epsilon}(q)$. In a Wannier basis, $\mathbf{\epsilon}(q)$ acquires generally a complex matrix structure. However, as shown in Ref. \cite{WFCE_2015}, environmental screening affects only a single element of the dielectric matrix $\mathbf{\epsilon}(q)$ which turns out to be accurately describable by continuum medium electrostatics. WFCE essentially modifies this one ``quasi macroscopic'' matrix element. The full WFCE algorithm which is described in Ref. \cite{WFCE_2015}.


      
%
	

Fig. \ref{fig:WFCEscheme}b demonstrates the range of applicability of the WFCE approach by comparing partially screened Coulomb interaction matrix elements in different graphene based systems obtained from direct cRPA calculations to WFCE results. In the latter case, an ab-initio calculation is necessary only for the bulk form of graphite. The WFCE approach predicts effective Coulomb matrix elements for monolayer and bilayer graphene very accurately; for instance, the
local Hubbard interaction agrees with the full cRPA calculation within $0.3$~eV. Larger deviations are only found in the case of graphene embedded in Ir metal, where hybridization of graphene and the surrounding material is sizable. I.e., the comparisons of full first-principles and WFCE calculations suggest that the WFCE approach is accurate when hybridization between layers in the vertical direction is not too strong, as is the case for van der Waals bonded systems.

Additionally, Fig. \ref{fig:WFCEscheme}b shows that Coulomb interactions in 2D materials like
graphene can be manipulated on the eV-scale by means of screening provided by different environments
which can be substrates, adsorbates, or other 2D materials. I.e. environmental screening presents a very effective tool for non-invasive materials manipulation. Possible applications can be the creation of heterojunctions by non-local manipulations of Coulomb interactions as put forward in Ref.~\cite{rosner_two-dimensional_2016}. However, this concept of ``Coulomb engineering'' is more general and can be likely also applied in the domain of strongly correlated systems like Mott insulators.

\subsection{Dynamical vertex approximation in nanosystems}
\label{sec:dga}

In low-dimensional systems, spatial correlations beyond mean-field become relevant. 
Non-local correlations have been taken into account following different strategies, 
ranging from quantum cluster theories,\cite{maierRMP77}  
which treat short-range correlations within the cluster exactly 
and long-range correlations within mean-field, 
to diagrammatic approximations, which treat correlations on all length-scales 
on equal footing, such as the dynamical vertex approximation
\cite{toschiPRB75,kataninPRB80,rohringerPRL107,schaeferPRB91,valliPRL104,valliPRB91} (D$\Gamma$A), 
the dual-fermion\cite{robtsovPRB77,rubtsovPRB79,hafermannPRL102,antipovPRL112} 
and one-particle irreducible\cite{rohringerPRB88} approaches,
as well as the DMF$^2$RG,\cite{tarantoPRL112} TRILEX,\cite{ayralPRB92,ayralPRB93} and QUADRILEX\cite{ayralPRB94}
In the following we discuss the idea behind the D$\Gamma$A 
and we present the first application of D$\Gamma$A in its parquet implementation, 
to nanoscopic systems.\cite{valliPRB91}

\begin{figure}
\resizebox{0.99\columnwidth}{!}{%
  \includegraphics{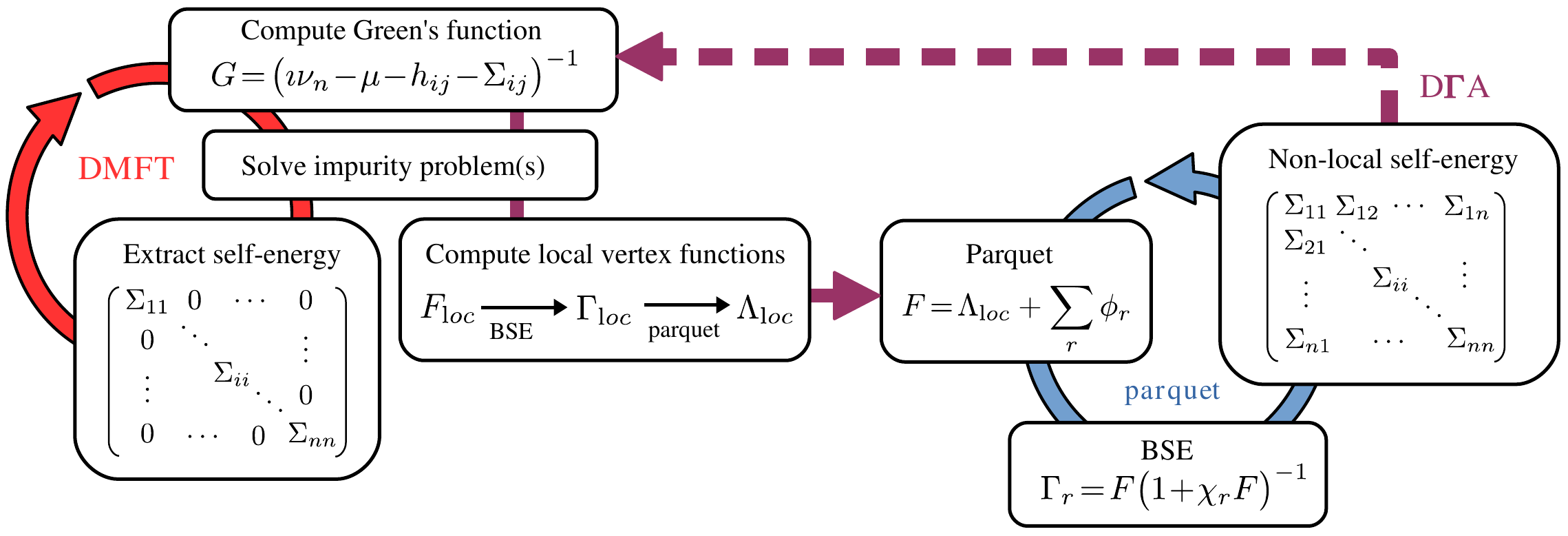} }
\caption{The flowchart compares the DMFT and D$\Gamma$A approaches. 
Non-local correlations can be included for the local fully-irreducible vertex $\Lambda_{\text loc}$ of a suitable AIM. 
The self-consistent solution the parquet equations with $\Lambda=\Lambda_{\textrm loc}$ 
allows to determine all the $k$-dependent vertex functions as well as the non-local self-energy. }
\label{fig:flowchart}
\end{figure}

From a diagrammatic prospective, the DMFT can be formulated in terms 
of a local approximation for the one-particle fully irreducible self-energy, 
owing to the locality of perturbation theory in infinite dimensions.\cite{georgesRMP68} 
DMFT has been successfully applied to model\cite{valliPRB86,valliPRB94,valliPRL104} 
as well as realistic\cite{jacobPRL103,Jacob_Kotliar_PRB2010,valliPRB92,dasPRL107} nanostructures.   
The idea behind DMFT can be systematically extended to include non-local corrections 
to the self-energy beyond mean-field, 
by considering diagrammatic approximations based on two-particle vertex functions. 
The Dyson-Schwinger equation  
\begin{equation}\label{eq:dyson-schwinger}
 \Sigma_{\bf k}(\nu) =     U \frac{n}{2} 
                        - T^2 U \sum_{\bf k'q}\sum_{\nu'\omega} 
                          F^{\uparrow\downarrow}_{\bf kk'q}(\nu,\nu',\omega)
                          G_{\bf k'+q}(\nu'+\omega) G_{\bf k'}(\nu') G_{\bf k+q}(\nu+\omega),
\end{equation}
relates the momentum-dependent electronic self-energy $\Sigma_{\bf k}(\nu)$ 
to the full two-particle vertex $F=F^{\uparrow\downarrow}_{\bf kk'q}(\nu,\nu',\omega)$.  
From a physical prospective, $F$ encodes all information 
of the two-body scattering processes in all particle-hole and particle-particle channels. 
Within the parquet formalism,\cite{bickers} $F$ can be decomposed 
according to \emph{reducibility} properties of two-particle diagrams as 
$F = \Lambda + \sum_r \phi_{r}$, where $\phi_r$ is the set of diagrams reducible 
in a specific scattering channel $r=ph,\overline{ph},pp$, 
while $\Lambda$ is the set of fully-irreducible diagrams (i.e., irreducible in all channels). 
Moreover, in each channel a Bethe-Salpeter equation (BSE) 
$F\!=\!\Gamma_r\!+\!\Gamma_r \chi_r F$ 
expresses $F$ in terms of the vertex $\Gamma_r$ , irreducible in channel $r$, 
in terms of the corresponding reducible susceptibility $\chi_r$. 

Each diagrammatic approach relies on a specific approximation 
at different levels of the diagrammatic theory.\cite{rohringerPRB86} 
The D$\Gamma$A assumes the locality of the fully-irreducible vertex,\cite{toschiPRB75} 
$\Lambda = \Lambda_{\textrm loc}$. 
Provided the knowledge of $\Lambda_{\textrm loc}$, the set of parquet equations above 
can be solved self-consistently to determine 
all the momentum-dependent vertex functions as well as the self-energy. 
Within the D$\Gamma$A $\Lambda_{\textrm loc}$ 
is evaluated from the auxiliary AIM, 
by solving the \emph{local} (inverse) parquet equation and the BSEs. 
A natural choice for the AIM is the (self-consistent) AIM of DMFT, 
which is expected to provide a reasonable description of local correlations 
of the original many-body problem.\cite{valliPRB91}
Otherwise, a fully self-consistent D$\Gamma$A calculation 
requires the AIM (and the corresponding $\Lambda_{\textrm loc}$) 
to be updated at each iteration until convergence. 
The full parquet-D$\Gamma$A scheme is shown in Fig.~\ref{fig:flowchart}.

We discuss the application of the D$\Gamma$A 
to cyclic organic molecules.\cite{valliPRL104,valliPRB91} 
Specifically, we consider the cyclic organic molecule [$n$]-annulene: C$_n$H$_n$. 
A proper description of electronic correlations is relevant 
for a proper estimate of the spectral gap, 
which deeply affects the electronic and transport properties of molecular junctions.  
We consider a low-energy effective model for the C$2p_z$ electrons, 
which are delocalized in a single bonding $\pi$-molecular orbital.
The Hamiltonian reads as: $H = H_{\textrm TB} + H_{\textrm int} + H_{\textrm hyb}$, 
where, $H_{\textrm TB}$ includes a nearest-neighbor tight-binding (or H{\"u}ckel) 
description of the molecule,  
$H_{\textrm int}$ is the local Coulomb interaction $U$ on the $\pi$-molecular orbital,  
and $H_{\textrm hyb}$ describes the (local) hybridization of each C atom 
to weakly-correlated electrodes or a substrate. 
For the sake of simplicity, we neglect non-local interactions. 
However, due to poor screening and quantum confinement effects, 
those may be expected to play an important role in determining the physics of nano systems, 
and can be taken into account within the D$\Gamma$A, 
as discussed e.g., in Refs.~\cite{toschiAP532} and \cite{galler1610.02998}. 

\begin{figure}
\resizebox{1.00\columnwidth}{!}{%
  \includegraphics{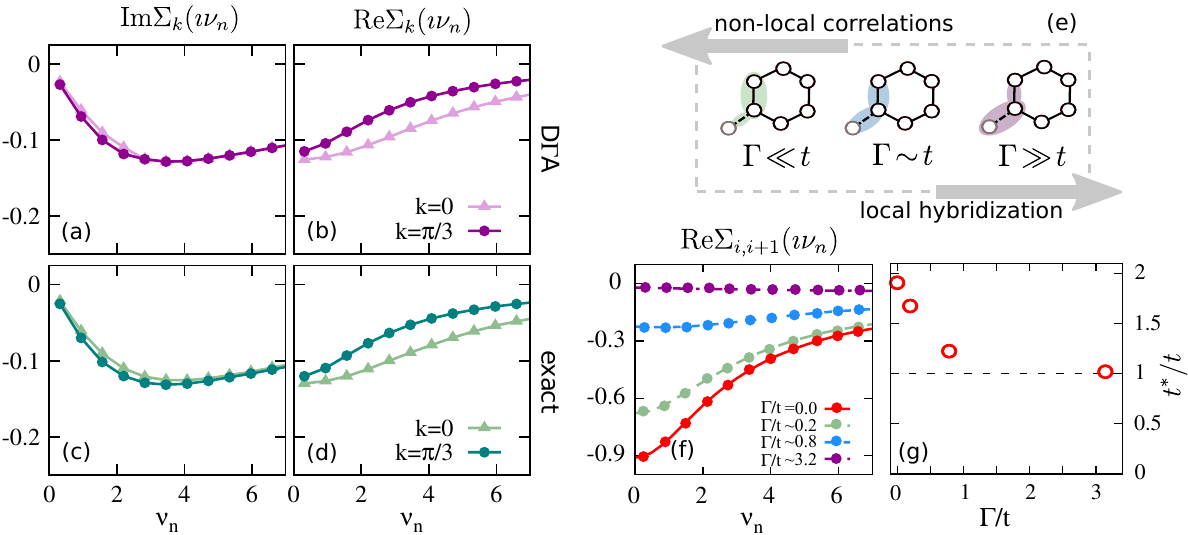} }
\caption{D$\Gamma$A (a-c) and exact (c-d) Matsubara self-energy 
for representative momenta $k\!=\!0$ and $k\!=\!\pi/3$ 
for the isolated benzene. 
The k-dependence of the scattering rate $\gamma$ is weak 
due to the presence of a spectral gap. 
Parameters: $U\!=\!2t$, $T\!=\!0.1t$. 
Real-space nearest-neighbor exact self-energy (f) and 
renormalization of the nearest-neighbor hopping $t^{*}/t$ (g) 
for different hybridization $\Gamma/t$. 
Parameters: $U\!=\!5t$, $T\!=\!0.05t$.
The interplay between local hybridization 
and non-local correlations is depicted in panel (e). 
Adapted with permission from Refs.~\cite{valliPRB86}~and~\cite{valliPRB91}. 
Copyrighted by the American Physical Society. }
\label{fig:siwijk_nc6}
\end{figure}

\begin{figure}\sidecaption
\resizebox{0.50\columnwidth}{!}{%
  \includegraphics{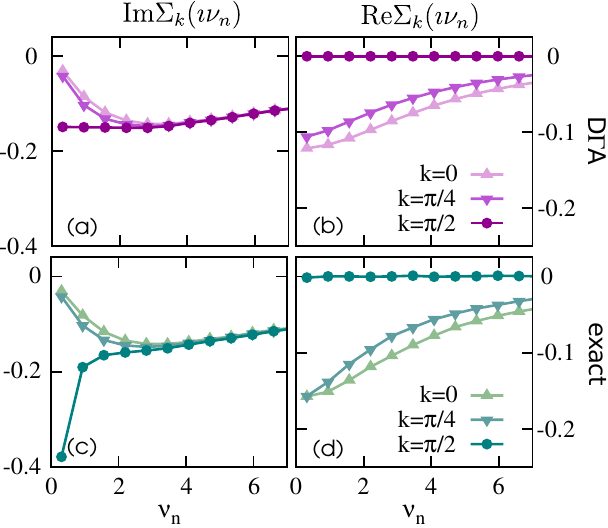} }
\caption{D$\Gamma$A (a-c) and exact (c-d) Matsubara self-energy 
for representative momenta $k\!=\!0$ and $k\!=\!\pi/4$ and $k\!=\!\pi/2$ 
for the isolated [8]-annulene.  
The D$\Gamma$A description is in overall good agreement with the exact solution, 
but for the large scattering rate at the Fermi surface $\gamma_{k\!=\!\pi/2}$ 
(see the text for a discussion). 
Parameters: $U\!=\!2t$, $T\!=\!0.1t$. 
Adapted with permission from Ref.~\cite{valliPRB91}. 
Copyrighted by the American Physical Society.}
\label{fig:siwk_nc6}
\end{figure}

Let us start considering the [6]-annulene (benzene). 
At half-filling, benzene is characterized by a semiconducting gap $\Delta_0\!=\!2|t|$ 
between the HOMO and LUMO molecular orbitals already in nearest-neighbor tight-binding model, 
in the absence of interaction. 
However, electronic correlations due to the local Coulomb repulsion renormalize the gap, 
deeply influencing the transport properties of benzene molecular junctions. 
It is known that local electronic correlations within DMFT 
\emph{reduce} the spectral gap in both bulk crystal\cite{sentefPRB80} 
and nanoscopic systems\cite{valliPRB86,valliPRB94} 
while the corresponding exact QMC solution yields $\Delta>\Delta_0$, 
suggesting an important role is played by non-local spatial correlations beyond DMFT. 
The D$\Gamma$A is able to capture all the relevant corrections beyond DMFT, 
and reproduces the exact self-energy, as shown in Fig.~\ref{fig:siwijk_nc6}(a-d). 
The scattering rate $\gamma_k\!\approx\!-2\imag\Sigma_{k}(\imath\nu_n\!\rightarrow\!0)$ 
is weakly $k$-dependent (and well described already within DMFT) 
due to the semiconducting nature of the benzene molecule in the gas phase. 
Instead, the strongly $k$-selective nature of the real part of the self-energy $\real\Sigma_{k}(\imath\nu_n)$, 
as shown both by the D$\Gamma$A and exact QMC solution, 
cannot be captured by DMFT, as $\real\Sigma(\imath\nu_n)\!=\!0$ for the local self-energy 
due to the particle-hole symmetry at half-filling.  
A real-space analysis\cite{valliPRB86,valliPhD} in Fig.~\ref{fig:siwijk_nc6}(f-g) shows that 
the nearest-neighbor hopping parameter is renormalized by interactions 
$t^{*}\!=\!t\!+\!\Sigma_{i,i\!+\!1}(\imath\nu_n\!\rightarrow\!0)$. 
In the presence of a local hybridization between each C atom in the molecule and the environment, 
described in the momentum-independent parameter $\Gamma$, 
non-local correlations are strongly suppressed. 
This determines, as a function of the control parameter $\Gamma/t$ a crossover 
between a weak-hybridization regime, where non-local correlations are dominant, 
to a local Kondo regime, characterized by the formation of entangled singlet states 
between the $\pi$-electron of benzene and the electrons of the environment, 
as shown in Fig.~\ref{fig:siwijk_nc6}(e).  
In the weak-hybridization limit, $\Delta\!\approx\!2|t^{*}|$ is a reasonable estimate for the spectral gap, 
in good agreement with the results obtained from the analytic continuation of the local Green's function.

\begin{figure}\sidecaption
\resizebox{0.55\columnwidth}{!}{%
  \includegraphics{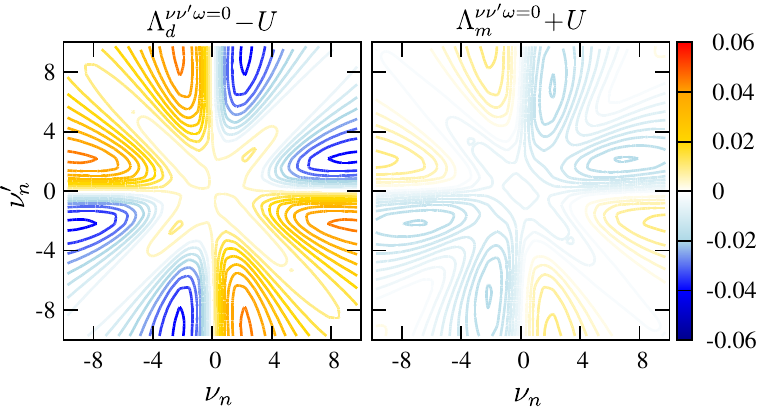} }
\caption{Typical ``butterly'' frequency structure of the local 
fully-irreducible vertex at the $\omega\!=\!0$ bosonic Matsubara frequency. 
$\Lambda_d$ and $\Lambda_m$ correspond to the vertex functions 
in the density and magnetic scattering channels, 
obtained from a self-consistent DMFT calculation for the isolated benzene molecule.  
Parameters: $U\!=\!2t$, $T\!=\!0.1t$.
Adapted with permission from Ref.~\cite{valliPRB91}. 
Copyrighted by the American Physical Society.} 
\label{fig:fir_nc6}
\end{figure}

\begin{figure}\sidecaption
\resizebox{0.55\columnwidth}{!}{%
  \includegraphics{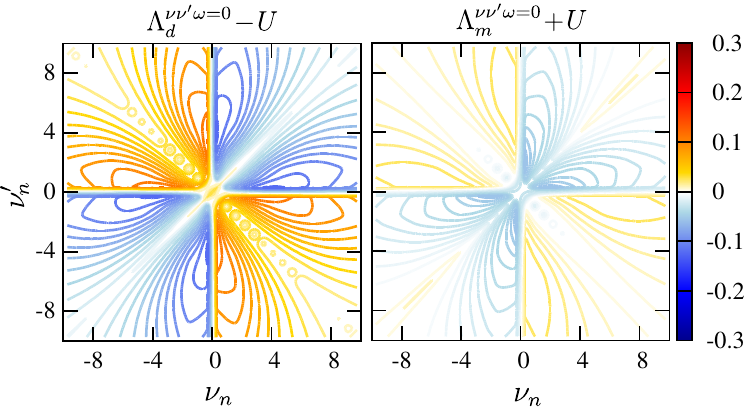} }
\caption{As for Fig.~\ref{fig:fir_nc6}, but obtained for the [8]-annulene. 
The low-energy frequency structure of $\Lambda_{d,m}$ reflects 
low-energy excitations due to the absence of a spectral gap within DMFT. 
Adapted with permission from Ref.~\cite{valliPRB91}. 
Copyrighted by the American Physical Society.}
\label{fig:fir_nc8}
\end{figure}

The situation is substantially different for the case 
of the half-filled isolated [8]-annulene. 
In the absence of interaction, $h_{ij}$ has a 2-fold degenerate eigenvalue 
at the Fermi level for $k=\pm\pi/2$(in contrast to the benzene molecule, which is semiconducting). 
Within DMFT the local spectral function displays a resonance at the Fermi level 
and undergoes a Mott metal-to-insulator transition at $U\!\approx\!8t$, 
while the exact solution displays a gap at any finite interaction\cite{valliPhD}. 
This is reflected in the large scattering rate $\gamma^{\textrm exact}_{k}$ 
(while $\real{\Sigma_k}\!=\!0$) for $k\!=\!\pi/2$. 
Instead, the D$\Gamma$A fails to open a gap at the Fermi level, 
as $\gamma^{\textrm D\Gamma A}_{k}\!\ll\!\gamma^{\textrm exact}_{k}$, 
resulting in a \emph{pseudogap}, associated with a partial transfer  
of the spectral weight toward higher energy.

The D$\Gamma$A results discussed above can be understood also from the analysis 
of the fully-irreducible vertex $\Lambda_{\textrm loc}$, 
which s shown in Figs.~\ref{fig:fir_nc6} and~\ref{fig:fir_nc8} 
in the density and magnetic (particle-hole) scattering channels. 
The frequency dependence of $\Lambda_{\textrm loc}$ 
has been discussed in detail if Ref.~\cite{rohringerPRB86} 
within the context of DMFT. 
Remarkably, the generic key features of $\Lambda_{\textrm loc}$, 
as the typical ``butterfly'' frequency structure of $\Lambda_{\textrm loc}^{\nu\nu'\omega=0}$, 
remain unchanged from the 2D Hubbard model to 0D nano systems. 
Compared to the fully-irreducible vertex of Ref.~\cite{rohringerPRB86} 
we notice that for the benzene both in $\Lambda_d$ and $\Lambda_m$ 
the low-energy features are substantially suppressed, in contrast to the case of the [8]-annulene, 
and similarly also for the [4]-annulene.\cite{valliPRB91}  
We ascribe this to the presence of a spectral gap 
in the local density of states in benzene, but not in the [4,8]-annulenes, within DMFT. 
This suggests that the difference between the D$\Gamma$A and exact solution in the case of the [8]-annulene, 
results from a poor descriptions of local correlations within DMFT. 
In the respect, a fully-self-consistent D$\Gamma$A calculations, 
although computationally expensive, it is expected 
to further improve the description of the low-energy physical properties of the [4,8]-annulenes.

%
%

\section{Conclusions and Outlook}


In this paper we have summarized the results of a series of collaborations between various groups at the University of W\"urzburg, the University of Bremen, the University of Hamburg and the TU Wien in Austria. One of the aims of this cooperative effort was to develop and optimize an interface between Kohn-Sham-based DFT and nano-DMFT/nano-D$\Gamma$A and achieve general applicability of these methods to complex correlated structures. Particular attention has been devoted to the choice of the basis sets and the influence of the double counting. 
These steps have been made possible thanks to the parallel development of our code package ``w2dynamics'', developed in W\"urzburg, Bremen as well as in Vienna\cite{parragh_2012,PhysRevB.92.155102,PhysRevB.94.125153}. ``w2dynamics'', which will be made public in the course of 2017 \cite{w2dynPAPER}, is a CT-QMC hybridization-expansion code, as described briefly in Sec.~\ref{sec:impsolv} and it allows users to set up calculations for big and complex realistic nanostructures in a rather flexible way. It is interfaced to the DFT-based packages used in the calculations shown here and it has been designed to be accessible not only to the experts of CT-QMC.
This is not the only advance in numerical algorithms for strongly correlated electron systems that we discuss in this review paper. As described in Sec. \ref{sec:varED}, we implemented a variational version of exact diagonalization \cite{PhysRevB.91.235142} as well as the computation of core level x-ray absorption and core level photo electron spectroscopy \cite{0295-5075-108-5-57004} in the framework of a new-generation Hamiltonian-based solver for DMFT \cite{PhysRevB.90.085102}. This is then extended to the case of multiorbital physics in nanostructures, which we approach with our combined CT-QMC and ED impurity solvers in order to establishing links from realistic structures to simple models.

Selected applications of the formalism in action have been presented in Sec. \ref{sec:applications}. In particular, as described in Sec. \ref{subsec:Hund}, some of us introduced the notion of Hund's impurities by studying Fe and Fe-hydrogen complexes adsorbed on a Pt(111) surface \cite{HundImp_2015} in collaboration with experimental colleagues from Hamburg. The role of the Hund's coupling in the Anderson model in general is also investigated, by comparing a multitude of calculations based on ED as well as CTQMC.
Furthermore, we have discussed the Kondo effect in TMPc molecules on Ag(001) as well as the effects of dehydrogenation seen in STM in Sec. \ref{subsec:molecules}. These systems allow for a direct experimental manipulation of the Kondo scale and offer a rich playground for the exploration of connections between theoretical models and real-world physics. 
Section \ref{subsec:Cr001} is instead devoted to the study on the Cr(001) surface \cite{schuler_many-body_2016}, which has been raising a long debate on the many-body nature of its low-energy spectral features. We reveal that the spectral feature observed close to the Fermi level is a consequence of changes in the electronic structure at the surface, as well as complex many-body effects in certain orbitals. We have been mostly concerned with further pursuing the realistic description of correlated impurities in environments with a reduced symmetry as well as of extended multi-orbital systems.
A related example, on which we however did not elaborate here, is the possibility of realizing a Mott transistor with $t_{2g}$-electrons in the context of oxide heterostructures, put forward in Ref.~\cite{PhysRevLett.114.246401}.

In this review, special attention has been given to non-local correlation effects and spatial correlations in nanoscopic systems, both touched upon in Sec. \ref{sec:non_loc}. We point out that in complex nanoscopic systems the role of non-local Coulomb interactions is important. In cooperation with the Research Center J\"ulich, we developed the Wannier function continuum electrostatics approach, showcased in Sec. \ref{subsec:wannier_electro}, which facilitates the realistic calculation of appropriately screened Coloumb interaction matrices in complex systems such as 2d materials supported by substrates \cite{WFCE_2015}. Generally low dimensional materials feature strong non-local Coulomb interaction which presents a big challenge for correlated materials modelling \cite{schuler_optimal_2013}. This issue has been addressed in Ref. \cite{van_loon_capturing_2016}: we showed that it is possible to link systems with non-local interactions to Hubbard models with local interactions only based on a thermodynamic variational principle. A further important worked accomplished in this context is Ref. \cite{PhysRevB.92.205132}.  

Specific applications of diagrammatic approaches to include non-local correlations in nanoscopic systems 
have been discussed in Sec. \ref{sec:dga}, dedicated among other things, 
to transport properties of nanoscopic systems, 
including single molecules \cite{valliPRL104,valliPRB86,valliPRB91} and quantum junctions \cite{valliPRL104,valliPRB86}.
In addition, some of us have analyzed in detail the frequency structure of the two-particle vertex \cite{PhysRevLett.110.246405,PhysRevLett.114.236402,PhysRevB.93.245102,PhysRevB.94.235108} and the role of spatial correlations \cite{valliPRB91,valliPRB92,rotterEPJB86} for several models of interacting as well as disordered electrons. We have reached a new level of understanding of the properties of diagrammatic expansions for many-particle systems. In particular, we have discovered fermionic divergencies of specific two-particle irreducible vertex functions and linked them with the appearance of ``protected'' degeneracies between physical and unphysical solutions of the Dyson equation.

The topics that our review has dealt with witness the fact that more and more complex systems became accessible to realistic dynamical mean-field theory simulations. I.e. a reliable description of local correlation effects is well doable now for many solid state systems. Yet, there are clearly challenges such as non-local correlation effects, competing interactions or also non-equilibrium effects in strongly correlated electron systems where progress has been made but where applications to complex real materials are at best at the very beginning to date and more development work is necessary.

\section*{Acknowledgments}
We gratefully acknowledge support of the Deutsche Forschungsgemeinschaft through FOR 1346. 
One of us (A.V.) acknowledges financial support from the Austrian Science Fund (FWF) 
through the Erwin Schr\"{o}dinger Fellowship No. J3890-N36. 

\bibliographystyle{epjbbib}
\bibliography{BibliogrFULL}


\end{document}